\documentclass[aps,prd,10pt,notitlepage,nofootinbib]{revtex4-1}

\usepackage[utf8]{inputenc}
\usepackage{amsmath,amssymb,amsfonts}
\usepackage{newtxtext,newtxmath}

\usepackage{bbold}
\usepackage{bm}
\usepackage{graphicx}
\usepackage[usenames,dvipsnames]{xcolor}
\usepackage{color}
\usepackage[colorlinks=true,linkcolor=blue,urlcolor=blue,citecolor=blue]{hyperref}

\usepackage{slashed}
\usepackage[english]{babel}
\usepackage{dcolumn}
\usepackage{pifont}
\usepackage{dsfont,mathrsfs}
\usepackage{cancel}
\usepackage{bigints}
\usepackage{accents}
\usepackage{soul}
\usepackage{multirow}
\usepackage{natbib}
\usepackage{tikz-feynman}

\usepackage{amsmath, amssymb}
\usepackage{bbold}
\usepackage{makecell}
\usepackage{simpler-wick}
\usepackage[export]{adjustbox}

\usepackage[bb=boondox]{mathalpha}

\usepackage{orcidlink} 

\newcommand{\fbr}[1]{\left(#1\right)}
\newcommand{\sbr}[1]{\left\{#1\right\}}
\newcommand{\tbr}[1]{\left[#1\right]}
\newcommand{\md}[1]{\left|#1\right|}
\newcommand{\mdvec}[1]{|{\vec{#1}}|}

\newcommand{\gm}{\gamma}

\newcommand{\n}{\nu}
\newcommand{\munu}{{\mu\nu}}
\newcommand{\alp}{\alpha}
\newcommand{\bt}{\beta}
\newcommand{\ep}{\epsilon}

\newcommand{\sK}{\slashed{K}}

\newcommand{\msd}{\mathscr{D}}
\newcommand{\mcpp}{\mathcal{P}_+}
\newcommand{\mcpm}{\mathcal{P}_-}

\newcommand{\tilpp}{\tilde{P}_+}
\newcommand{\tilpm}{\tilde{P}_-}
\newcommand{\tilkp}{\tilde{K}_+}
\newcommand{\tilkm}{\tilde{K}_-}
\newcommand{\tilskm}{\tilde{\sK}_-}
\newcommand{\tilskp}{\tilde{\sK}_+}

\newcommand{\util}{\tilde{U}}
\newcommand{\gtil}{\tilde{g}}

\newcommand{\wkr}{\omega_k^r}
\newcommand{\wpr}{\omega_p^r}
\newcommand{\wkp}{\omega_k^+}
\newcommand{\wkm}{\omega_k^-}
\newcommand{\wpp}{\omega_p^+}
\newcommand{\wpm}{\omega_p^-}

\newcommand{\fppp}{f_+^{p+}}
\newcommand{\fmpp}{f_-^{p+}}
\newcommand{\fppm}{f_+^{p-}}
\newcommand{\fmpm}{f_-^{p-}}
\newcommand{\fpkp}{f_+^{k+}}
\newcommand{\fmkp}{f_-^{k+}}
\newcommand{\fpkm}{f_+^{k-}}
\newcommand{\fmkm}{f_-^{k-}}

\newcommand{\MD}{M_D}

\newcommand{\logterm}{\log{\frac{q_0+q}{q_0-q}}}

\newcommand{\nn}{\nonumber}
\newcommand{\Sll}[1]{S_{11}\fbr{#1}}

\newcommand{\IM}{\text{Im~}}
\newcommand{\RE}{\text{Re~}}

\newcommand{\ensembleaverage}[1]{\left\langle#1\right\rangle}

\newcommand{\FB}[1]{(#1)}
\newcommand{\SB}[1]{\left\{#1\right\}}
\newcommand{\TB}[1]{\left[#1\right]}

\newcommand{\mcTc}{\mathcal{T}_C}

\newcommand{\llp}{{l l^\prime}}

\newcommand{\Ja}{J_\alpha}
\newcommand{\Jbd}{J_\beta^\dagger}

\newcommand{\Dab}{\bm{ D}_{\alpha \beta}}

\newcommand{\DR}{D^R_{\alpha \beta}}

\newcommand{\SpF}{\rho_{\alpha \beta}}

\newcommand{\DG}{D^>_{\alpha \beta}}
\newcommand{\DL}{D^<_{\alpha \beta}}

\newcommand{\Xint}{\int d^4 X}

\newcommand{\Pint}{\int \dfrac{d^4 P}{(2\pi)^4}}

\newcommand{\FT}[2]{e^{i #1 \cdot #2}}
\newcommand{\IFT}[2]{e^{-i #1 \cdot #2}}

\allowdisplaybreaks

\begin{document}
	\title{Collective phenomena in chirally imbalanced medium}
			
	\author{Sourav Duari\orcidlink{0009-0006-0795-5186}$^{a,c}$}
	\email{s.duari@vecc.gov.in}
	\email{sduari.vecc@gmail.com}
	
	\author{Nilanjan Chaudhuri\orcidlink{0000-0002-7776-3503}$^{a,c}$}
	\email{n.chaudhri@vecc.gov.in}
	\email{nilanjan.vecc@gmail.com}

	\author{Sourav Sarkar\orcidlink{0000-0002-2952-3767}$^{a,c}$}
	\email{sourav@vecc.gov.in}
	
	\author{Pradip Roy\orcidlink{0009-0002-7233-4408}$^{b,c}$}
	\email{pradipk.roy@saha.ac.in}

	\affiliation{$^a$Variable Energy Cyclotron Centre, 1/AF Bidhannagar, Kolkata - 700064, India}
	\affiliation{$^b$Saha Institute of Nuclear Physics, 1/AF Bidhannagar, Kolkata - 700064, India}
	\affiliation{$^c$Homi Bhabha National Institute, Training School Complex, Anushaktinagar, Mumbai - 400085, India}

	\begin{abstract}
		We calculate the gluon polarization tensor for a chirally imbalanced plasma  using hard thermal loop approximation in the real time formulation of thermal field theory. The dispersion relations obtained from the poles of the effective gluon propagator are solved numerically as well as analytically in appropriate limiting cases. It is seen that the degenerate transverse modes split into left and right handed circularly polarized modes. We also compute imaginary poles of the propagator which signal the presence of instability in the plasma. Relevant time scales for development of such instabilities are discussed in detail. Furthermore, we compute both the real and imaginary parts of the static heavy-quark potential in the chirally imbalanced plasma and argue that quarkonium suppression is enhanced due to the combined effects of a reduced debye screening length and an increased decay width. In addition, we calculate the gluon spectral density, sum rules and residues for various cases, providing a comprehensive understanding of the collective behaviour of the medium.

	\end{abstract}
	
	\maketitle
	\section{Introduction}
	\newcommand{\ignore}[1]{}

	The study of the vacuum structure of quantum chromodynamics (QCD) under extreme conditions of temperature and/or baryon density is a primary goal of relativistic heavy ion collision (HIC) experiments conducted at RHIC and LHC. 	It is well known that, at zero and low temperatures, QCD has an infinite number of energy-degenerate vacuum configurations which can be characterized by topologically non-trivial gauge fields with a non-zero winding number~\cite{Shifman:1988zk}. 
	These configurations, known as instantons, facilitate transitions between different vacua by overcoming a potential barrier whose height is of the order of the QCD scale $\Lambda_{\rm QCD}$. This process is  referred to as instanton tunneling and has been extensively studied in the literature~\cite{Belavin:1975fg,tHooft:1976rip,tHooft:1976snw}.
	However, at high temperatures, such as those achieved in the quark-gluon plasma (QGP) phase during HICs, another type of gluonic configuration, called sphalerons, is expected to be produced abundantly~\cite{Manton:1983nd,Klinkhamer:1984di}.
	It is hypothesized that the presence of sphalerons significantly enhances the transition rates across the barriers separating the energy-degenerate vacua~\cite{Kuzmin:1985mm,Arnold:1987mh,Khlebnikov:1988sr,Arnold:1987zg}.
	These topologically non-trivial gauge field configurations interact with quarks, enabling changes in their helicities. Such interactions lead to the local breaking of parity ($ P $) and charge-parity ($ CP $) symmetries by inducing an asymmetry between left- and right-handed quarks through the axial anomaly of QCD~\cite{Adler:1969gk, Bell:1969ts}.
	While no global $ P $ and $ CP $-violating effects have been directly observed in QCD~\cite{Adler:1969gk, Bell:1969ts, McLerran:1990de, Moore:2010jd}, a locally generated chirality imbalance may occur. 
	This local imbalance is described using the chiral chemical potential (CCP), which represents the difference in the number of right and left-handed quarks.
	The CCP is defined as $ \mu_5 = (\mu_R - \mu_L )/2 $, where $ \mu_L $
	and $ \mu_R $ denote the chemical potentials of the left and right-handed fermions respectively.
	Chiral systems hold significant relevance across various fields, including particle physics, nuclear physics, condensed matter physics, and cosmology. A notable example is the chiral magnetic effect (CME), which involves generation of a parity-violating current within a plasma when exposed to a magnetic field.
	The CME is of particular interest in the context of heavy-ion collisions. Despite substantial efforts by the STAR Collaboration to detect the CME at RHIC using isobaric collisions, no clear evidence has yet been observed~\cite{STAR:2021mii}. Consequently, new experimental techniques for the detection of the CME have been proposed making it an active area of ongoing research.~\cite{An:2021wof, Milton:2021wku}.

	The spectrum of collective excitations is a fundamental feature of any many-body system. It provides crucial insights into the thermodynamic and transport properties of systems in equilibrium and  largely governs the temporal evolution of systems out of equilibrium.
	When elementary particles propagate through a medium their properties change due to interactions and become `dressed'. This includes acquiring effective masses different from those in vacuum. More broadly, such propagation leads to  collective modes or quasi-particles which often have dispersion relations and behaviours distinct from those in vacuum. 
	In some cases there are also imaginary modes which lead to instability in the plasma helping in rapid thermalization~\cite{Mrowczynski:1993qm,Arnold:2003rq,Kurkela:2011ti,Kurkela:2011ub,Ipp:2010uy,Attems:2012js}. Collective modes in the plasma have been extensively studied using both kinetic theory and diagrammatic approach~\cite{Mrowczynski:2000ed,Mrowczynski:2004kv}. For example, in Ref.~\cite{Blaizot:1993zk,Kelly:1994ig,Litim:2001db} the equivalence of both these methods has been established for isotropic and chirally symmetric plasma. The same for anisotropic and chirally symmetric systems has been shown in Refs.~\cite{Mrowczynski:2000ed}. 

	The polarization tensor in a chirally asymmetric system is first calculated in~\cite{Son:2012zy} using the kinetic theory approach. These findings are then used in~\cite{Akamatsu:2013pjd} to demonstrate that even an isotropic plasma can sustain imaginary collective modes which leads to instabilities in the chiral plasma at finite temperature. The same approach is used in~\cite{Carignano:2018thu} to investigate collective modes in ultradegenerate chiral matter. Additionally, Ref.~\cite{Carrington:2021bnk} shows that presence of anisotropy significantly enhances the effect of the chiral chemical potential resulting in much larger imaginary modes compared to the isotropic case even for moderate levels of anisotropy.

	In this work we calculate the gluon self-energy in a chiral plasma at finite temperature in the real time formalism using the hard thermal loop (HTL) approximation. We explicitly show that the components of the polarization tensor match with those in Ref~\ignore{PRL111052002}\cite{Akamatsu:2013pjd} obtained in the kinetic theory approach. The dispersion relations are obtained from the poles of the effective propagator. 
	It is well established that due to the small relative velocities of heavy quarks, quarkonium systems can be effectively studied in the non-relativistic limit. As a result, several in-medium properties, such as the masses and decay rates of heavy-quark (HQ) bound states can be obtained by solving the quantum mechanical Schrödinger equation with a complex HQ potential. Recent progress in this area includes first-principles lattice QCD simulations that allow extraction of the complex HQ potential~\cite{Rothkopf:2011db,Burnier:2014ssa}, as well as the development of complex-valued potential models aimed at quantitatively capturing quarkonium behavior in a medium~\cite{Strickland:2011aa,Burnier:2015nsa,Guo:2018vwy}.	The HQ potential consists of a real part which determines the binding energy and an imaginary part arising from Landau damping of low-frequency gauge fields and color singlet–octet transitions. The imaginary part encodes information about quarkonium dissociation via wave-function decoherence~\cite{Laine:2006ns,Brambilla:2008cx,Rothkopf:2019ipj}. In this work, we also compute both the real and imaginary parts of the static heavy-quark potential in a chiral plasma at finite temperature.
	Additionally we evaluate the sum rules and residues which are of relevance for the calculations of energy loss, transport coefficients etc.

	This article is organized as follows. In Sec. II the calculation of gluon self energy in presence of chirally imbalanced medium is shown and the effective gluon propagator is obtained in Sec. III. Explicit evaluations of structure functions are presented in Sec. IV. Dispersion relations are shown in Sec.V and phenomena such as splitting of transverse modes into left- and right handed circularly polarized modes are discussed along with the imaginary modes which are associated with instability of the plasma. In Sec. VI we calculate the Debye radius and heavy quark potential in chirally asymmetric medium. In Sec VII sum rules and residues for spectral density are calculated. Sec VIII provides the summary and conclusions. Some aspects of the real-time formalism used in this work are provided in the Appendix.


\section{One-loop self-energy of gluons }

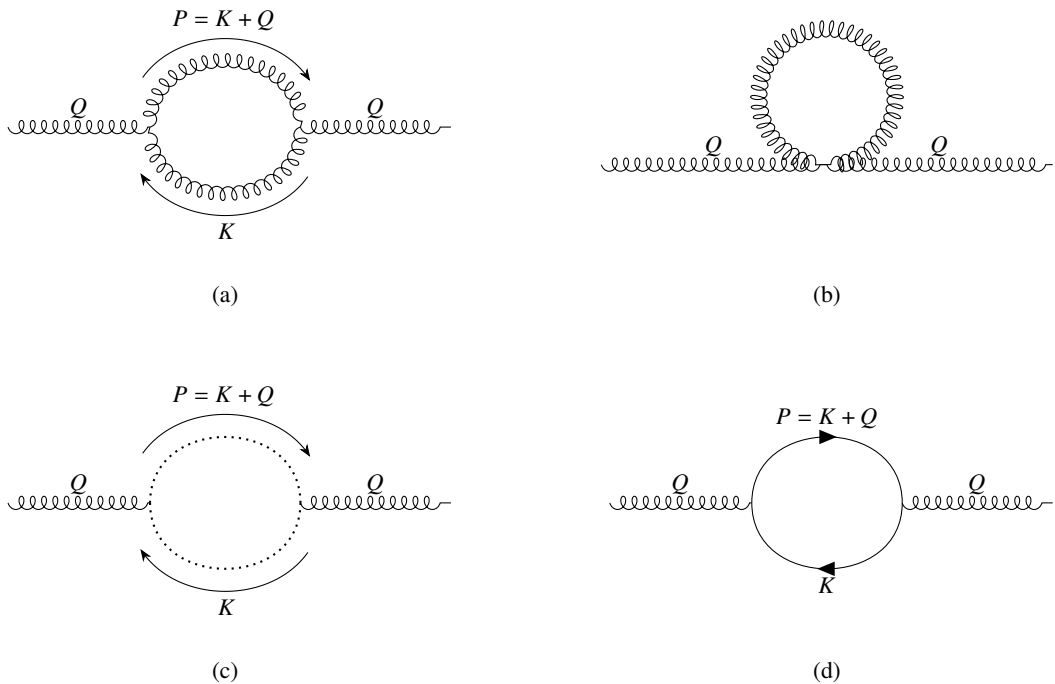
\begin{figure}[h]
	
	\begin{tikzpicture}	
		
		\begin{feynman} 
			
			\vertex (a1) {};
			\vertex[right=0cm of a1] (a1) {};
			\vertex [right=2cm of a1] (b1);
			\vertex[right=2cm of b1] (c1);
			\vertex[right=2 cm of c1] (d1);
			\vertex[right=1 cm of b1] (p);
			\vertex[below=2 cm of p] (p){(a)};

			\diagram* {
				(a1) -- [gluon,edge label=\(Q\)] (b1),
				(c1) -- [gluon,edge label=\(Q\)] (d1),
				(c1) -- [gluon, half left, momentum=\(K\)] (b1),
				(b1) -- [gluon, half left, momentum=\({P=K+Q}\)] (c1),
			};

		\end{feynman}

		\begin{feynman} 
			
			\vertex[right=8cm of a1] (a2) {};
			\vertex[below=0.5cm of a2] (a2);
			\vertex [right=3cm of a2] (b2);
			\vertex[right=3cm of b2] (c2);
			\vertex[above=2cm of b2] (d2);
			\vertex[below=1.5 cm of b2] (p){(b)};

			\diagram* {
				(a2) -- [gluon,edge label=\(Q\)] (b2),
				(b2) -- [gluon,edge label=\(Q\)] (c2),
			};	
			
			\draw[
			black,
			decorate,
			decoration={
				coil,
				segment length=3.5pt,
				amplitude=3pt,
				post length=0.5mm,
				pre length=0.5mm
			},
			] (b2) arc [start angle=-90, end angle=270, radius=0.9cm];

		\end{feynman}

		\begin{feynman} 
			
			\vertex[below=5cm of a1] (a3) {};
			\vertex [right=2cm of a3] (b3);
			\vertex[right=2cm of b3] (c3);
			\vertex[right=2 cm of c3] (d3);
			\vertex[right=1 cm of b3] (p);
			\vertex[below=2 cm of p] (p){(c)};

			\diagram* {
				(a3) -- [gluon,edge label=\(Q\)] (b3),
				(c3) -- [gluon,edge label=\(Q\)] (d3),
				(c3) -- [ghost, half left, momentum=\(K\)] (b3),
				(b3) -- [ghost, half left, momentum=\({P=K+Q}\)] (c3),
			};

		\end{feynman}

		\begin{feynman} 
			
			\vertex[right=8cm of a3] (a1) {};
			\vertex [right=2cm of a1] (b1);
			\vertex[right=2cm of b1] (c1);
			\vertex[right=2 cm of c1] (d1);
			\vertex[right=1 cm of b1] (p);
			\vertex[below=2 cm of p] (p){(d)};

			\diagram* {
				(a1) -- [gluon,edge label=\(Q\)] (b1),
				(c1) -- [gluon,edge label=\(Q\)] (d1),
				(c1) -- [fermion, half left, edge label=\(K\)] (b1),
				(b1) -- [fermion, half left, edge label=\({P=K+Q}\)] (c1),
			};

		\end{feynman}	
		
	\end{tikzpicture}
	\caption{Feynman diagrams for one-loop gluon self-energy}\label{one-loop}
\end{figure}

The Feynman diagrams contributing to the gluon self-energy are shown in Fig.~\ref{one-loop} of which the first three diagrams do not depend on the chiral chemical potential and their contributions are well-known. So in the gluon self energy chiral asymmetry enters only in Fig.~\ref{one-loop}(d) where only quarks are involved. In this paper, we will use real time formulation (RTF) of thermal field theory where all two-point correlation functions such as the propagator and self-energy become $2\times2$ matrices in the thermal space~\cite{Bellac:2011kqa,Mallik:2016anp}. Few key features of RTF are summarized in Appendix~\ref{App_RTF}. Here we calculate the 11-component of the gluon polarization tensor for Fig.~\ref{one-loop}(d) which is necessary to study collective properties of the chiral plasma~\cite{Weldon:1982aq}. It is given by
\begin{equation}\label{self energy}
	 \Pi_{11}^{\mu\n}=-ig^2\frac{N_f}{2}\int \frac{d^4K}{(2\pi)^4}\text{Tr}\tbr{\gm^\mu\Sll{P=K+Q}\gm^\n\Sll{K}}
\end{equation}
where $g$ is the strong coupling constant, $N_f$ is the number of quark flavours, $Q=(q^0,\vec q)$ is the external gluon momentum and $K=(k^0,\vec k)$ is the loop momentum. $\Sll{K}$ is the 11-component of the fermionic thermal propagator matrix~\cite{Ghosh:2022xbf} in presence of chiral imbalanced medium in RTF and is given by
\begin{equation}
	\Sll{K}=\msd(k^0+\mu,\mdvec{k}) \sum_{r\in \sbr{\pm}}\frac{1}{4\mdvec{k}r\mu_5}\tbr{\frac{-1}{\fbr{k^0+\mu}^2-(\wkr)^2+i\ep}-\eta(k^0+\mu)2\pi i\delta((k^0+\mu)^2-(\wkr)^2)}
\end{equation}
Here $\eta(x)=\Theta(x)f_+(x)+\Theta(-x)f_-(-x)$, $f_\pm(x)=\frac{1}{e^{\frac{x\mp\mu}{T}}+1}$, $\wpr=\mdvec{p}+r\mu_5$, $\wkr=\mdvec{k}+r\mu_5$.  $\msd(k^0+\mu,\mdvec{k})$ and $\msd(p^0+\mu,\mdvec{p})$ contain the Dirac structure of the fermion propagators which are given by
\begin{equation}
	\msd(k^0+\mu,\mdvec{k})= \mcpp \tilkm^2 \tilskp+\mcpm \tilkp^2 \tilskm 
\end{equation}
in which $\mathcal{P}_\pm = \frac{1}{2}(1\pm\gm^5)$, $\tilde{K}_\pm^\mu\equiv(k^0+\mu\pm\mu_5,\vec{k})$ and  $\tilde{P}_\pm^\mu\equiv(p^0+\mu\pm\mu_5,\vec{p})$.
Eq.~\eqref{self energy} can be written as
\begin{equation}
	\Pi^\munu_{11}=\fbr{\Pi^\munu_{11}}_{\rm vac}+\fbr{\Pi^\munu_{11}}_I+\fbr{\Pi^\munu_{11}}_{II},
\end{equation}
where $\fbr{\Pi^\munu_{11}}_{\rm vac}$ is the temperature independent part and $\fbr{\Pi^\munu_{11}}_I$, $\fbr{\Pi^\munu_{11}}_{II}$ are temperature dependent parts. These are given by 
\begin{eqnarray}\label{self_energy_vac}
	\fbr{\Pi^\munu_{11}}_{\rm vac}&=&-\frac{ig^2}{(2\pi)^4}\frac{N_f}{2}\int d^4K \frac{L^\munu}{16 kp\mu_5}\tbr{\frac{-1}{(k^0+\mu)^2-(\wkp)^2+i\epsilon}-\frac{-1}{(k^0+\mu)^2-(\wkm)^2+i\epsilon}}\nn\\
	&&\tbr{\frac{-1}{(p^0+\mu)^2-(\wpp)^2+i\epsilon}-\frac{-1}{(p^0+\mu)^2-(\wpm)^2+i\epsilon}}
\end{eqnarray}
\begin{eqnarray}\label{self_energy_I}
	\fbr{\Pi^\munu_{11}}_I&=&-\frac{ig^2}{(2\pi)^4}\frac{N_f}{2}\int d^4K\frac{L^\munu}{16kp\mu_5^2}\tbr{2\pi i \eta(p^0+\mu)\sbr{\frac{1}{(k^0+\mu)^2-(\wkp)^2+i\epsilon} -\frac{1}{(k^0+\mu)^2-(\wkm)^2+i\epsilon} }\right.\nn\\
	&&\sbr{\delta\fbr{(p^0+\mu)^2-(\wpp)^2}-\delta\fbr{(p^0+\mu)^2-(\wpm)^2}}
 +2\pi i \eta(k^0+\mu)\sbr{\frac{1}{(p^0+\mu)^2-(\wpp)^2+i\epsilon} -\frac{1}{(p^0+\mu)^2-(\wpm)^2+i\epsilon} }\nn\\
&&\left.\sbr{\delta\fbr{(k^0+\mu)^2-(\wpp)^2}-\delta\fbr{(k^0+\mu)^2-(\wpm)^2}}}
\end{eqnarray}
\begin{eqnarray}\label{self_energy_II}
	\fbr{\Pi^\munu_{11}}_{II}&=&-\frac{ig^2}{(2\pi)^4}\frac{N_f}{2}\int d^4K\frac{L^\munu}{16kp\mu_5^2}(2\pi i \eta(k^0+\mu))(2\pi i \eta(p^0+\mu))\sbr{\delta\fbr{(k^0+\mu)^2-(\wkp)^2}-\delta\fbr{(k^0+\mu)^2-(\wkm)^2}}\nn\\
	&&\sbr{\delta\fbr{(p^0+\mu)^2-(\wpp)^2}-\delta\fbr{(p^0+\mu)^2-(\wpm)^2}}~.
\end{eqnarray}
Here $k=\mdvec{k}$, $p=\mdvec{p}$ and
\begin{eqnarray}
 L^\munu&=&2\tilpm^2\tilkm^2\tbr{\tilkp^\mu\tilpp^\nu+\tilkp^\nu\tilpp^\mu-g^\munu\tilkp\cdot \tilpp}+2i\tilpm^2\tilkm^2 \epsilon^{\mu\nu\alpha\beta} (\tilpp)_\alpha (\tilkp)_\beta+\nn\\
 &&2\tilpp^2\tilkp^2\tbr{\tilkm^\mu\tilpm^\nu+\tilkm^\nu\tilpm^\mu-g^\munu\tilkm\cdot \tilpm}-2i\tilpp^2\tilkp^2 \epsilon^{\mu\nu\alpha\beta} (\tilpm)_\alpha (\tilkm)_\beta~.
\end{eqnarray}
Only the real part of the analytic in-medium self energy i.e. the diagonal component of $2\times 2$ matrix in the thermal space contributes to the dispersion relations~\cite{Weldon:1982aq}. This can be calculated from the 11-component of the self energy by using Eq.~\eqref{Eq_Re11_Re}. Note that, in the appendix we have distinguished the diagonal element from other components by denoting it via a \textit{`bar'}. However in the following discussion only the diagonal component is necessary to calculate different collective properties of the system and hence for clarity of presentation we will remove the \textit{`bar'} notation. Since Eq.~\eqref{self_energy_II} only contributes to the imaginary part,  integrating over $k^0$ in Eq.~\eqref{self_energy_I} we get
\begin{eqnarray}\label{Re-self energy}
	\text{Re}\Pi^\munu&=&\frac{g^2}{(2\pi)^3}\frac{N_f}{2}\int\frac{d^3k}{16kp\mu_5^2}\text{P}\left[\fppp \frac{L_5^\munu}{2\wpp}\left\{\frac{1}{2\wkp}\fbr{\frac{1}{-q^0-\wkp+\wpp}-\frac{1}{-q^0+\wkp+\wpp}}\right.\right.\nn\\
	&&\left.-\frac{1}{2\wkm}\fbr{\frac{1}{-q^0-\wkp+\wpp}-\frac{1}{-q^0+\wkp+\wpp}}\right\}-\fppm\frac{L_5^{\prime\munu}}{2\wpm}\left\{\frac{1}{2\wkp}\fbr{\frac{1}{-q^0-\wkp+\wpm}-\frac{1}{-q^0+\wkp+\wpm}}\right.\nn\\
	&&\left.-\frac{1}{2\wkm}\fbr{\frac{1}{-q^0-\wkm+\wpm}-\frac{1}{-q^0+\wkm+\wpm}}\right\}+\fmpp\frac{L_6^{\munu}}{2\wpp}\left\{\frac{1}{2\wkp}\fbr{\frac{1}{-q^0-\wkp-\wpp}-\frac{1}{-q^0+\wkp-\wpp}}\right.\nn\\
	&&\left.-\frac{1}{2\wkm}\fbr{\frac{1}{-q^0-\wkm-\wpp}-\frac{1}{-q^0+\wkm-\wpp}}\right\}-\fmpm\frac{L_6^{\prime\munu}}{2\wpm}\left\{\frac{1}{2\wkp}\fbr{\frac{1}{-q^0-\wkp-\wpm}-\frac{1}{-q^0+\wkp-\wpm}}\right.\nn\\
	&&\left.-\frac{1}{2\wkm}\fbr{\frac{1}{-q^0-\wkm-\wpm}-\frac{1}{-q^0+\wkm-\wpm}}\right\}+\fppm\frac{L_6^{\prime\munu}}{2\wpm}\left\{\frac{1}{2\wkp}\fbr{\frac{1}{-q^0-\wkp-\wpm}-\frac{1}{-q^0+\wkp-\wpm}}\right.\nn\\
	&&\left.-\frac{1}{2\wkm}\fbr{\frac{1}{-q^0-\wkm-\wpm}-\frac{1}{-q^0+\wkm-\wpm}}\right\}+\fpkp\frac{L_1^{\munu}}{2\wkp}\left\{\frac{1}{2\wpp}\fbr{\frac{1}{q^0+\wkp-\wpp}-\frac{1}{q^0+\wkp+\wpp}}\right.\nn\\
	&&\left.-\frac{1}{2\wpm}\fbr{\frac{1}{q^0+\wkp-\wpm}-\frac{1}{q^0+\wkp+\wpm}}\right\}-\fpkm\frac{L_1^{\prime\munu}}{2\wkm}\left\{\frac{1}{2\wpp}\fbr{\frac{1}{q^0+\wkm-\wpp}-\frac{1}{q^0+\wkm+\wpp}}\right.\nn\\
	&&\left.-\frac{1}{2\wpm}\fbr{\frac{1}{q^0+\wkm-\wpm}-\frac{1}{q^0+\wkm+\wpm}}\right\}+\fmkp\frac{L_2^{\munu}}{2\wkp}\left\{\frac{1}{2\wpp}\fbr{\frac{1}{q^0-\wkp-\wpp}-\frac{1}{q^0-\wkp+\wpp}}\right.\nn\\
	&&\left.-\frac{1}{2\wpm}\fbr{\frac{1}{q^0-\wkp-\wpm}-\frac{1}{q^0-\wkp+\wpm}}\right\}-\fmkm\frac{L_2^{\prime\munu}}{2\wkm}\left\{\frac{1}{2\wpp}\fbr{\frac{1}{q^0-\wkm-\wpp}-\frac{1}{q^0-\wkm+\wpp}}\right.\nn\\
	&&\left.\left.-\frac{1}{2\wpm}\fbr{\frac{1}{q^0-\wkm-\wpm}-\frac{1}{q^0-\wkm+\wpm}}\right\}\right]
\end{eqnarray}
where $L_1^{\munu}=L^{\munu}{\big{\arrowvert_{k_0=-\mu+\omega_k^+}}}$, $L_1^{\prime\munu}=L^{\munu}{\big{\arrowvert_{k_0=-\mu+\omega_k^-}}}$, $L_2^{\munu}=L^{\munu}{\big{\arrowvert_{k_0=-\mu-\omega_k^+}}}$, $L_2^{\prime\munu}=L^{\munu}{\big{\arrowvert_{k_0=-\mu-\omega_k^-}}}$, $L_5^{\munu}=L^{\munu}{\big{\arrowvert_{k_0=-q_0-\mu+\omega_p^+}}}$, $L_5^{\prime\munu}=L^{\munu}{\big{\arrowvert_{k_0=-q_0-\mu+\omega_p^-}}}$, $L_6^{\munu}=L^{\munu}{\big{\arrowvert_{k_0=-q_0-\mu-\omega_p^+}}}$, $L_6^{\prime\munu}=L^{\munu}{\big{\arrowvert_{k_0=-q_0-\mu-\omega_p^-}}}$, $f_+^{k+}=f_+(\omega_k^+) $, $f_-^{k+}=f_-(\omega_k^+) $, $f_+^{k-}=f_+(\omega_k^-) $, $f_-^{k-}=f_-(\omega_k^-) $, $f_+^{p+}=f_+(-q_0+\omega_p^+) $, $f_-^{p+}=f_-(q_0+\omega_p^+) $, $f_+^{p-}=f_+(-q_0+\omega_p^-) $ and $f_-^{p-}=f_-(q_0+\omega_p^-) $.

\section{ General structure of gluon self-energy and the effective propagator}

The general structure of gluon self-energy in a chirally imbalanced medium is given by\cite{Nieves:1988qz}
\begin{equation}\label{Eq_Gen_SE}
	\Pi^\munu = \Pi_T R^\munu_T + \Pi_L Q^\munu_L   + \Pi_A P^\munu_A~,
\end{equation}
where  $Q^\munu_L$, $R_T^\munu$ and $P_A^\munu$ are orthogonal to each other and also orthogonal to $Q_\mu$ and $Q_\nu$ so that gauge invariance is preserved i.e $Q_\mu\Pi^\munu=Q_\nu\Pi^\munu=0$. $Q_L^\munu$, $R_T^\munu$ and $P_A^\munu$ are constructed from the known quantities i.e gluon momentum $q^\mu$ , metric tensor $g^\munu$ and thermal bath velocity $U^\mu$. Form of these tensors are as follows,
\begin{equation}
	Q^\munu_L=\frac{\util^\mu\util^\nu}{\util^2},\hspace{0.5cm}R_T^\munu=\gtil^\munu-Q^\munu_L,\hspace{0.5cm} P_A^\munu=\frac{i}{q}\ep^{\munu\alp\bt}Q_\alp U_\bt~,
\end{equation}
where 
\begin{equation}
	\util^\mu=\gtil^\munu U_\nu,\hspace{0.5cm}\gtil^\munu=g^\munu-\frac{Q^\mu Q^\nu}{Q^2},
\end{equation}
and $U^\mu=(1,0,0,0)$ in the local rest frame.
$\Pi_L$ and $\Pi_T$ are longitudinal and transverse parts of gluon polarization tensor and $\Pi_A$ is the anomalous  contribution which appears due to the chiral asymmetry of the medium.
\begin{equation}\label{coefficient}
	\Pi_L=(Q_L)_\munu\Pi^\munu,\hspace{0.5cm}\Pi_T=\frac{1}{2}(R_T)_\munu\Pi^\munu,\hspace{0.5cm}\Pi_A=-\frac{1}{2}(P_A)_\munu\Pi^\munu,\hspace{0.5cm}
\end{equation}


\begin{figure}[h]
	\centerline{\includegraphics[scale=.30]{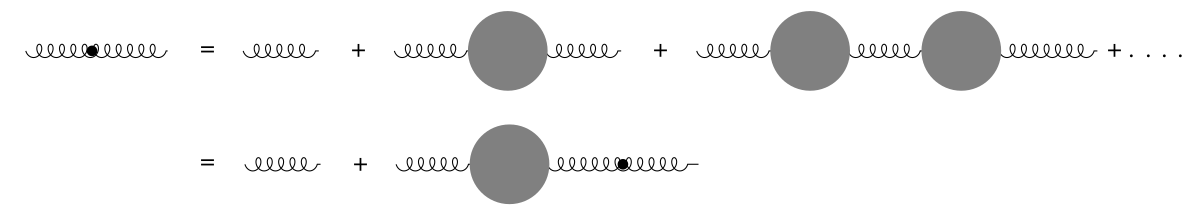}}
	\caption{Effective gluon propagator}
	\label{dyson_fig}
\end{figure}
Using the Dyson-Schwinger equation (Fig.~\ref{dyson_fig}) we get
\begin{equation}
	(G^{-1})_\munu=(G^{-1}_0)_\munu+\Pi_\munu
\end{equation}  
where $(G^{-1}_0)_\munu$ is inverse of the free gluon propagator given by
\begin{equation}
	(G^{-1}_0)_\munu=Q^2g_\munu-\fbr{1-\frac{1}{\xi}}Q_\mu Q_\nu~.
\end{equation}
We now decompose $(G)_\munu$ as
\begin{equation}
	G_\munu=G_1 Q_\mu Q_\nu+G_2 A_\munu+ G_3 (Q_L)_\munu+G_4 B_\munu
\end{equation}
where $ A_\munu$, $ (Q_L)_\munu$ and $ B_\munu$ are the projection tensors defined as
\begin{equation}
	A_\munu=\frac{1}{2}\fbr{ (R_T)_\munu+ (P_A)_\munu},\hspace{0.5cm}  B_\munu=\frac{1}{2}\fbr{ (R_T)_\munu- (P_A)_\munu}~.
\end{equation}
The effective propagator is finally given by
\begin{equation}\label{Eff_prop}
	G_\munu=\frac{\xi}{Q^2}\frac{Q_\mu Q_\nu}{Q^2}+\frac{A_\munu}{Q^2+\Pi_T+\Pi_A}+\frac{B_\munu}{Q^2+\Pi_T-\Pi_A}+\frac{(Q_L)_\munu}{Q^2+\Pi_L}~.
\end{equation}

\section{Evaluation of the structure functions $\Pi_L$, $\Pi_T$, $\Pi_A$ }
\subsection{Calculation of $\Pi_L$}

Using the transversality condition i.e $q_\mu\Pi^\munu=0$ we can write 
\begin{equation}\label{eq_piL}
	\Pi_L=(Q_L)_\munu\Pi^\munu=\frac{U_\mu U_\nu}{\util^2}\Pi^\munu=\frac{\Pi^{00}}{\util^2}~.
\end{equation}
Now, in the HTL approximation the external soft momentum ($~gT$) can be ignored  with respect to the loop momentum which is of order $T$. This leads to the following approximate expressions
	\begin{eqnarray}\label{approx}
		p\approx k+q\cos\theta~~~~~~~~~~~~~~~~~~~\nn\\
		\frac{1}{\pm q_0\pm(\wpp-\wkp)}\approx\frac{1}{\pm q_0\pm q\cos\theta}~~~~~~~~~~\nn\\
		\frac{1}{\pm q_0\pm(\wpm-\wkm)}\approx\frac{1}{\pm q_0\pm q\cos\theta}~~~~~~~~~~\nn\\	
		\frac{1}{\pm q_0\pm(\wpp-\wkm)}\approx \frac{1}{\pm 2\mu_5\pm q^0\pm q\cos\theta}~~~~~~~\nn\\
		\frac{1}{\pm q_0\pm(\wpm-\wkp)}\approx\frac{1}{\mp 2\mu_5\pm q^0\pm q\cos\theta}~~~~~~~\nn\\
		\frac{1}{\pm q^0+\omega^\pm_p+\omega^\pm_k}\approx\frac{1}{2k\pm2\mu_5\pm q^0+q\cos\theta}~~~~~~\nn\\
		\frac{1}{\pm q^0+\omega^\pm_p+\omega^\mp_k}\approx\frac{1}{2k\pm q^0+q\cos\theta}~~~~~~~~\nn\\
		f_\pm^{p_\pm}\approx f_\pm^{k_\pm}+\frac{\partial f_\pm^{k_\pm}}{\partial k} q\cos\theta~~~~~~~~~~\nn\\
	\end{eqnarray}
	using which we can write
	\begin{equation}\label{piL-L1}
		L_1^{00}\approx L_2^{ 00}\approx L_5^{00} \approx L_6^{00} \approx 64\mu_5^2 k^2 (k+\mu_5)^2
	\end{equation}
	\begin{equation}\label{piL-L1p}
		L_1^{\prime 00}\approx L_2^{\prime 00}\approx L_5^{\prime00} \approx L_6^{\prime 00} \approx 64\mu_5^2 k^2 (k-\mu_5)^2~.
	\end{equation}
From Eqs.~\eqref{Re-self energy},~\eqref{eq_piL}-\eqref{piL-L1p} the
expression for $\Pi_L$ after some algebra becomes
\begin{eqnarray}
	\Pi_L&\approx&\frac{1}{\util^2}\frac{g^2}{(2\pi)^3}\frac{N_f}{2}\tbr{\int d^3k\fbr{\frac{\partial\fpkp}{\partial k}+\frac{\partial\fpkm}{\partial k}}\frac{q\cos\theta}{-q_0+q\cos\theta}+\int d^3k\fbr{\frac{\partial\fmkp}{\partial k}+\frac{\partial\fmkm}{\partial k}}\frac{q\cos\theta}{q_0+q\cos\theta}}\nn\\
	&=& g^2\frac{N_f}{2}\fbr{\frac{T^2}{3}+\frac{\mu^2+\mu_5^2}{\pi^2}}\frac{q_0^2-q^2}{q^2}\fbr{1-\frac{q_0}{2q}\logterm}~.
\end{eqnarray}
Now, the results of the loop diagrams in Fig.~\ref{one-loop}a-c are $\mu_5$ independent and exist in the literature~\cite{Bellac:2011kqa,Klimov:1982bv,Weldon:1982aq}. Considering all four loops of Fig.~\ref{one-loop} we can write $\Pi_L$ as
\begin{eqnarray}
	\Pi_L&=& \fbr{g^2\frac{T^2}{6}\fbr{N_f+2C_A}+\frac{N_fg^2}{2\pi^2}\fbr{\mu^2+\mu_5^2}} \frac{q_0^2-q^2}{q^2}\fbr{1-\frac{q_0}{2q}\logterm}\nn\\
	&=&\MD^2 \frac{q_0^2-q^2}{q^2}\fbr{1-\frac{q_0}{2q}\logterm}
\end{eqnarray}
where $\MD$ is the Debye mass given by
\begin{equation}
	\MD^2=g^2\frac{T^2}{6}\fbr{N_f+2C_A}+\frac{N_fg^2}{2\pi^2}\fbr{\mu^2+\mu_5^2}~.
\end{equation}


\subsection{Calculation of $\Pi_T$}

From Eq.~\eqref{coefficient} we can write
\begin{eqnarray}
	\Pi_T&=&\frac{1}{2}(R_T)_\munu\Pi^\munu=\frac{1}{2}\tbr{\gtil_\munu-(Q_L)_\munu}\Pi^\munu\nn\\
	&=&\frac{1}{2}\gtil_\munu\Pi^\munu-\frac{1}{2}\Pi_L~.
\end{eqnarray}
The transversality condition i.e $q_\mu\Pi^\munu=0$ gives
\begin{equation}\label{eq_piT}
	\Pi_T=\frac{1}{2}g_\munu\Pi^\munu-\frac{1}{2}\Pi_L~.
\end{equation}
Using  Eqs.~\eqref{approx} the relevant quantities can be written as
\begin{equation}\label{piT-L1}
	g_\munu L_1^{\munu}\approx -g_\munu L_5^{\munu}\approx -64 k\mu_5^2(k+\mu_5)^2(q_0-q\cos\theta)
\end{equation}
\begin{equation}\label{piT-L1p}
	g_\munu L_1^{\prime \munu}\approx -g_\munu L_5^{\prime \munu}\approx -64 k\mu_5^2(k-\mu_5)^2(q_0-q\cos\theta)
\end{equation}
\begin{equation}\label{piT-L2}
	g_\munu L_2^{\munu}\approx -g_\munu L_6^{\munu}\approx 64 k\mu_5^2(k+\mu_5)^2(q_0+q\cos\theta)
\end{equation}
\begin{equation}\label{piT-L2p}
	g_\munu L_2^{\prime \munu}\approx -g_\munu L_6^{\prime \munu}\approx 64 k\mu_5^2(k-\mu_5)^2(q_0+q\cos\theta)~.
\end{equation}
From Eqs.~\eqref{Re-self energy},\eqref{approx},\eqref{eq_piT}-\eqref{piT-L2p} we can write $\Pi_T$ as
\begin{eqnarray}
	\Pi_T&\approx&-\frac{g^2}{(2\pi)^3}\frac{N_f}{2}\int\frac{d^3k}{k}\tbr{\fpkp+\fmkm+\fmkp+\fpkm}-\frac{1}{2}\Pi_L\nn\\
	&=&-\frac{1}{2}g^2\frac{N_f}{2}\fbr{\frac{T^2}{3}+\frac{\mu^2+\mu_5^2}{\pi^2}}-\frac{1}{2}\Pi_L\nn\\
	&=&-\frac{1}{2}g^2\frac{N_f}{2}\fbr{\frac{T^2}{3}+\frac{\mu^2+\mu_5^2}{\pi^2}}\frac{q_0^2}{q^2}\fbr{1+\frac{1}{2}\fbr{\frac{q}{q_0}-\frac{q_0}{q}}\logterm}~.
\end{eqnarray}

Considering all four loops of Fig.~\ref{one-loop} the final expression for $\Pi_T$ is given by
\begin{eqnarray}
	\Pi_T &=& -\frac{1}{2}\fbr{g^2\frac{T^2}{6}\fbr{N_f+2C_A}+\frac{N_fg^2}{2\pi^2}\fbr{\mu^2+\mu_5^2}} \frac{q_0^2}{q^2}\fbr{1+\frac{1}{2}\fbr{\frac{q}{q_0}-\frac{q_0}{q}}\logterm}\nn\\
	&=&-\frac{1}{2}\MD^2 \frac{q_0^2}{q^2}\fbr{1+\frac{1}{2}\fbr{\frac{q}{q_0}-\frac{q_0}{q}}\logterm}~.
\end{eqnarray}

\subsection{Calculation of $\Pi_A$}
To calculate $\Pi_A$ we need to evaluate $L^\munu (P_A)_\munu$ where
\begin{equation}
	L^\munu (P_A)_\munu=4\frac{\tilkp^2\tilpp^2}{q}\tbr{\tilkm\cdot Q \tilpm^0-\tilpm\cdot Q \tilkm^0}+4\frac{\tilkm^2\tilpm^2}{q}\tbr{\tilpp\cdot Q \tilkp^0-\tilkp\cdot Q \tilpp^0}~.
\end{equation}
So the required quantities in the HTL approximation are given by
\begin{equation}\label{piP-L1}
	L_1^{\munu}(P_A)_\munu \approx L_5^{\munu}(P_A)_\munu \approx 64 k \mu_5^2(k+\mu_5)^2\frac{q_0}{q}\fbr{q_0-q\cos\theta-\frac{q_0^2-q^2}{q_0}}
\end{equation}
\begin{equation}\label{piP-L1p}
	L_1^{\prime\munu}(P_A)_\munu \approx L_5^{\prime\munu}(P_A)_\munu \approx -64 k \mu_5^2(k-\mu_5)^2\frac{q_0}{q}\fbr{q_0-q\cos\theta-\frac{q_0^2-q^2}{q_0}}
\end{equation}
\begin{equation}\label{piP-L2}
	L_2^{\munu}(P_A)_\munu \approx L_6^{\munu}(P_A)_\munu \approx 64 k \mu_5^2(k+\mu_5)^2\frac{q_0}{q}\fbr{q_0+q\cos\theta-\frac{q_0^2-q^2}{q_0}}
\end{equation}
\begin{equation}\label{piP-L2p}
	L_2^{\prime\munu}(P_A)_\munu \approx L_6^{\prime\munu}(P_A)_\munu \approx -64 k \mu_5^2(k-\mu_5)^2\frac{q_0}{q}\fbr{q_0+q\cos\theta-\frac{q_0^2-q^2}{q_0}}~.
\end{equation}
Using Eqs.~\eqref{Re-self energy},\eqref{approx},\eqref{piP-L1}-\eqref{piP-L2p} we can write $\Pi_A$ as
\begin{eqnarray}
	\Pi_A&\approx&\frac{g^2}{(2\pi)^2}\frac{N_f}{2}\frac{q_0^2-q^2}{q}\fbr{1-\frac{q_0}{2q}\logterm}\int dk k \tbr{\frac{\partial\fpkp}{\partial k}-\frac{\partial\fmkm}{\partial k}-\frac{\partial\fpkm}{\partial k}+\frac{\partial\fmkp}{\partial k}}\nn\\
	&=&-\frac{g^2}{2\pi^2}\frac{N_f}{2} \mu_5 \frac{q_0^2-q^2}{q}\fbr{1-\frac{q_0}{2q}\logterm}~.
\end{eqnarray}

	
\section{Collective Modes}

Both stable and unstable gluonic collective modes are present in the chiral plasma. 
These are obtained from the poles of the effective gluon propagator given by Eq.~\eqref{Eff_prop} which ultimately boils down to solving the following equations
\begin{align}
	Q^2+\Pi_L &= 0  \label{dis1} \\
	Q^2+\Pi_T\pm\Pi_A &= 0~. \label{dis2}
\end{align}
Eqs.~\eqref{dis1} and \eqref{dis2} describe the dispersion relations for longitudinal and transverse modes respectively.
From Eq.~\eqref{dis2} one can infer that the degeneracy of the transverse mode is lifted due to presence of chiral imbalance.
Stable modes correspond to real solutions of these dispersion relations while unstable modes arise from purely imaginary solutions.


\begin{figure}[h]
	\includegraphics[scale=0.375]{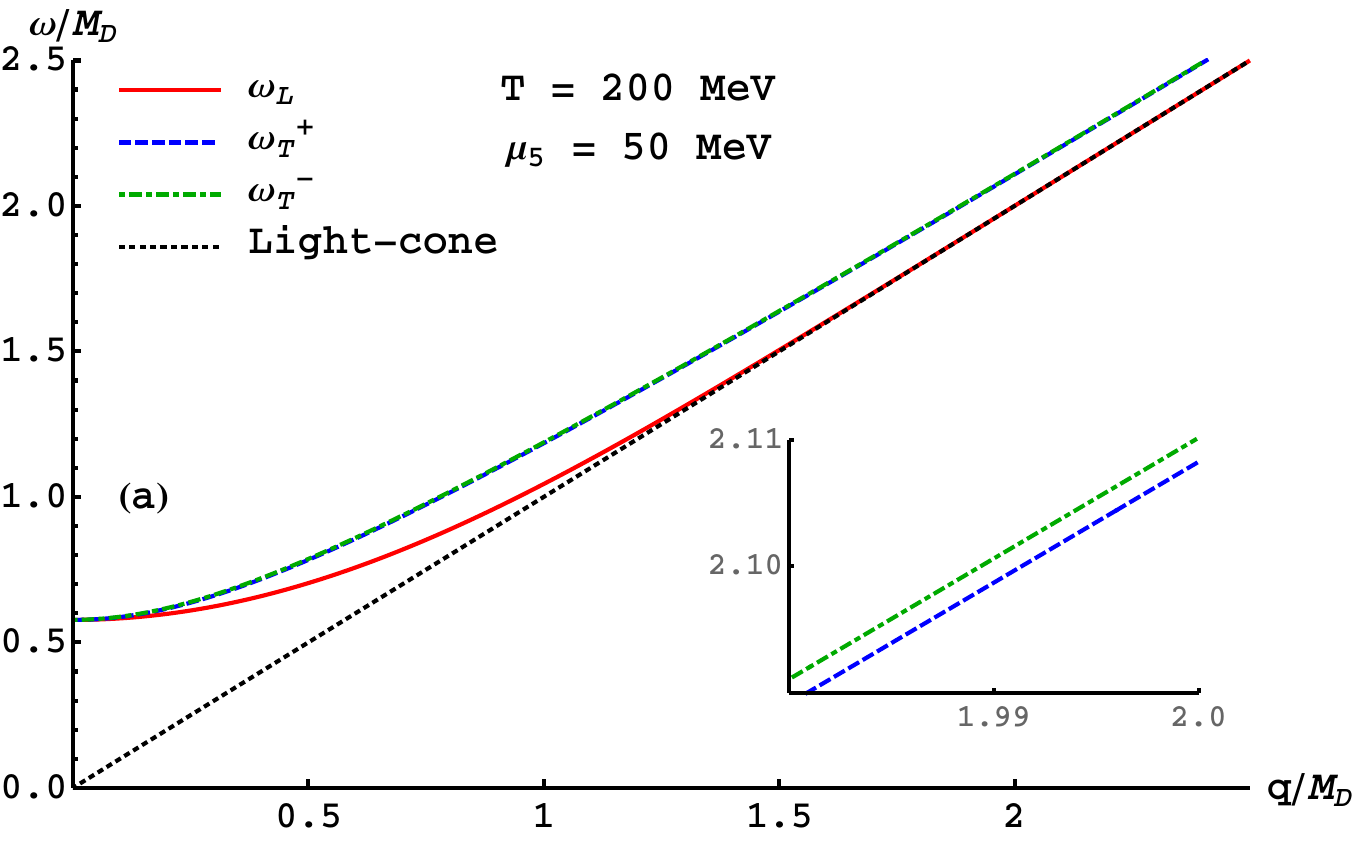} ~~~~
	\includegraphics[scale=0.375]{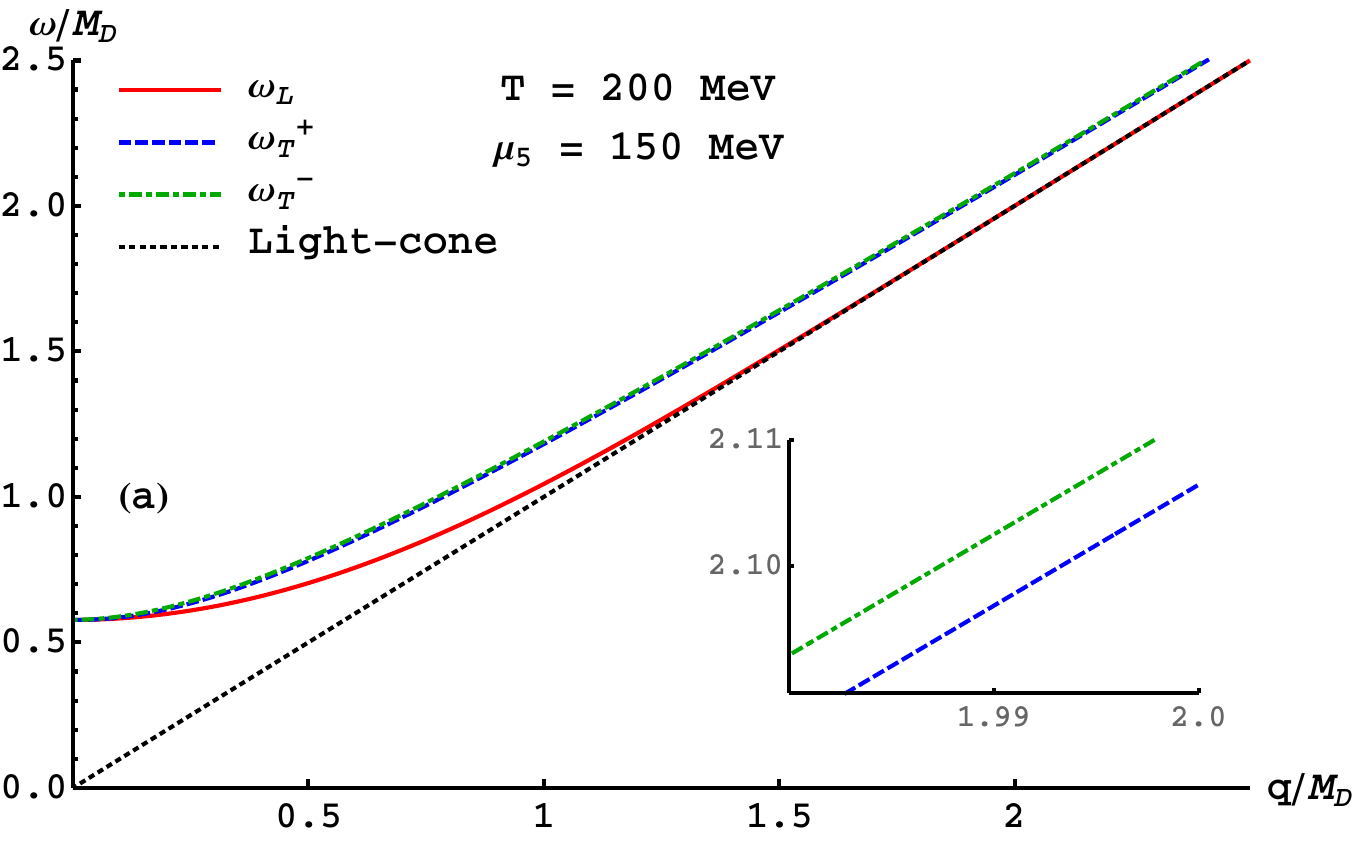} \\ \textcolor{white}{text} \vspace{1. em}\\
	\includegraphics[scale=0.375]{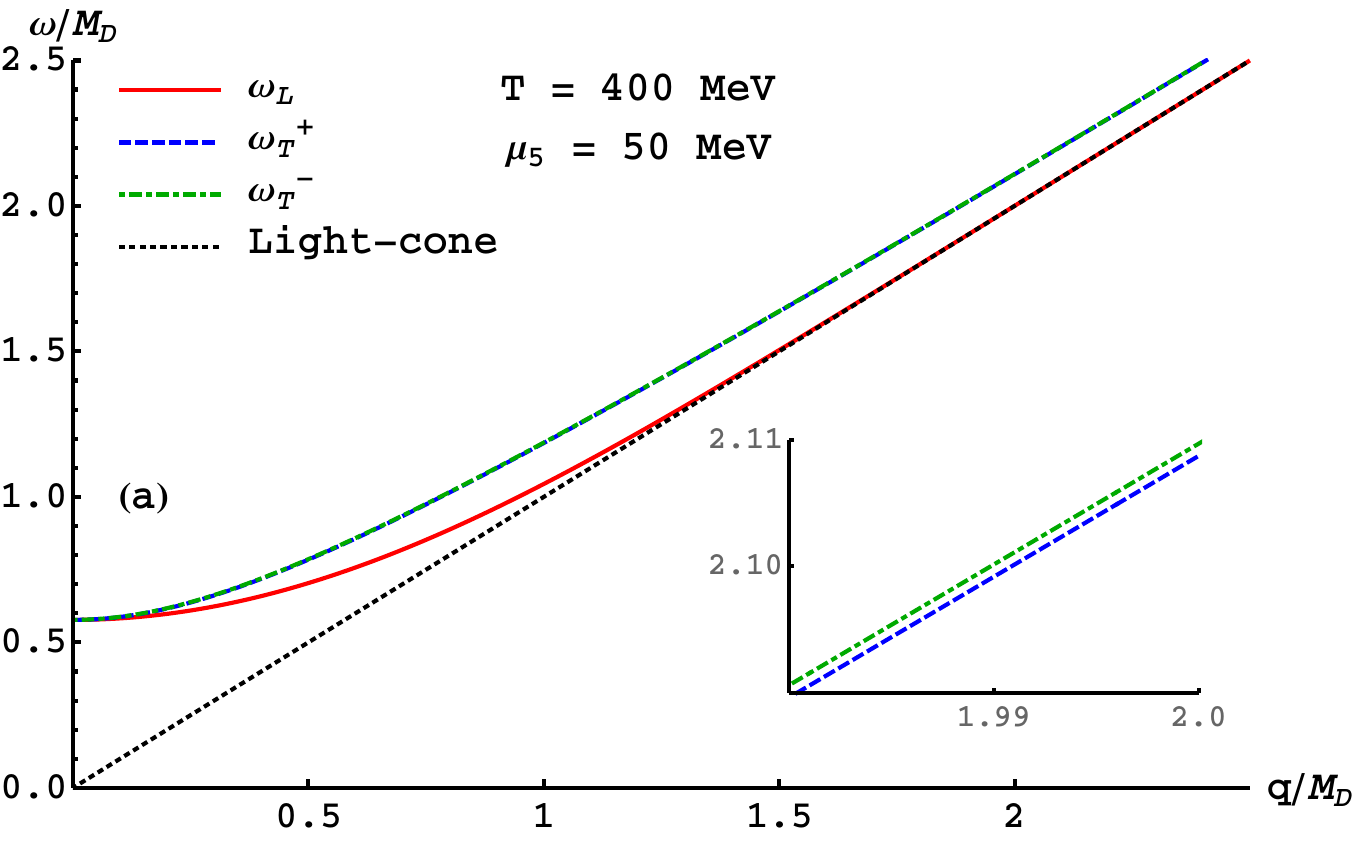} ~~~~
	\includegraphics[scale=0.375]{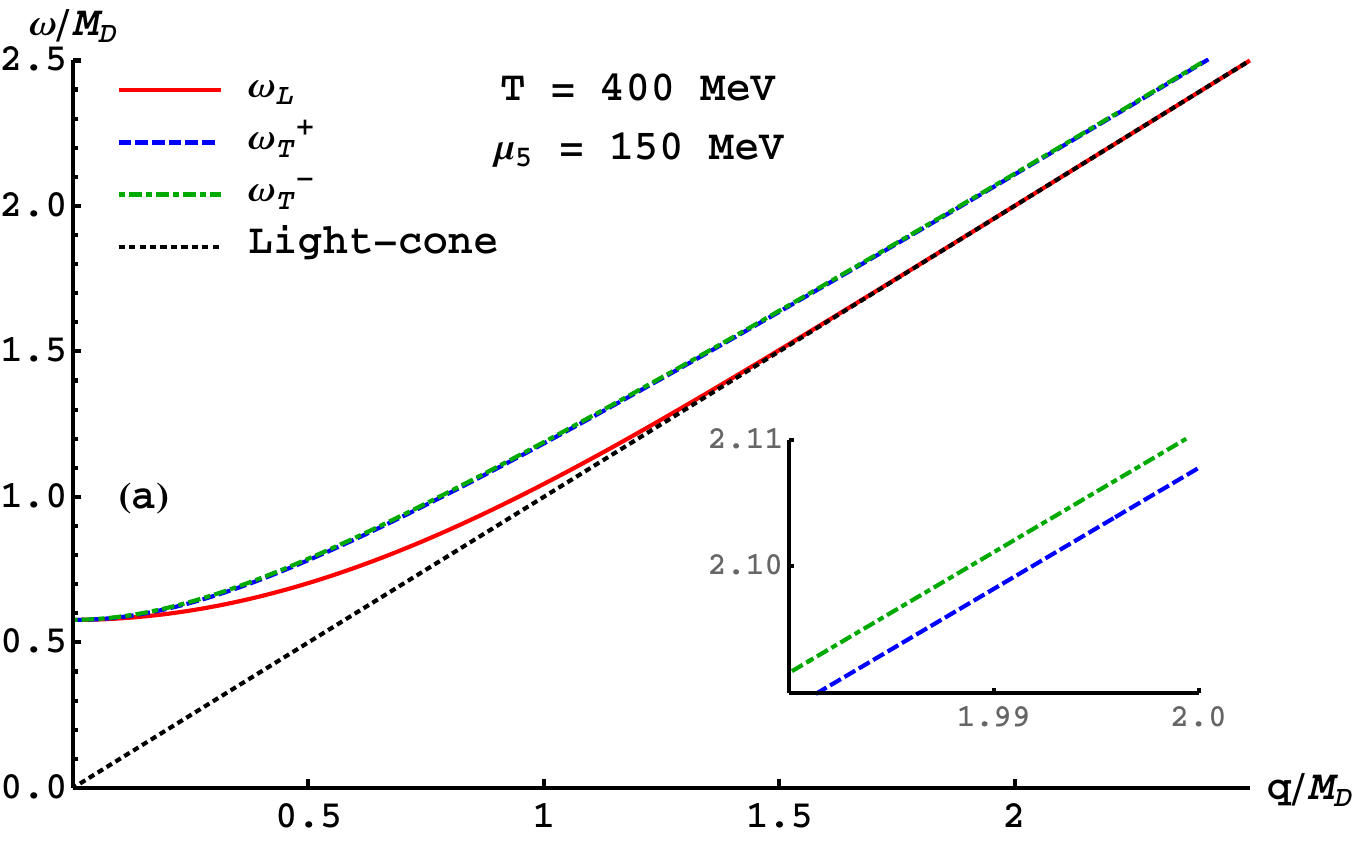} 

	\caption{Gluonic modes in hot chirally asymmetric medium }\label{wtl}
\end{figure}

In Figs.~\ref{wtl} (a)$ - $(d) we have presented longitudinal and transverse gluonic modes for different physical scenarios which correspond to the real solutions of Eqs.~\eqref{dis1} and \eqref{dis2}. Solution of Eq.~\eqref{dis2} with positive (negative) sign is denoted as $ \omega_T^+ $ ($ \omega_T^- $). To obtain these results we have considered a representative value of the quark chemical potential $ \mu = \mu_R + \mu_L = 150 $~MeV. Without loss of generality, we assume positive values for  $ \mu_5 $, which implies that the net number of right-handed quarks exceeds the net number of left-handed quarks in the system. Additionally two different values of temperature $ T = 200, 400 $~MeV and chiral chemical potential $ \mu_5  = 50, 150$~MeV are chosen which are relevant for the QGP phase. 
In all the plots the collective mode of the gluon splits into three distinct modes due to the presence of a medium with chiral imbalance. The long wavelength longitudinal mode is known as the `plasmon mode' (shown in red dash-dot line) which arises solely due to the presence of a thermal medium. In the large (hard) momentum limit it reduces to the free dispersion and decouples from the plasma. Two transverse modes, represented by solid blue and dashed green lines, are found to exhibit distinct dispersive behaviours (inset plots are included for clarity). This difference arises purely due to the presence of finite $ \mu_5 $ as the modes are otherwise doubly degenerate in a thermal medium~\cite{Bellac:2011kqa}. For higher values of momentum both of these modes resemble the transverse gluons in vacuum. Thus it can be inferred that at small momenta all the collective modes are equally important.
Furthermore, both the transverse and longitudinal modes lie above the light cone ($ \omega = q $) in all cases. As the dispersion curves remain above the light cone, the phase velocity of the plasma waves exceeds the speed of light. Consequently, no Landau damping occurs as this mechanism requires the velocity of a plasma particle to match the phase velocity of the wave. 
Now comparing Figs.~\ref{wtl} (a) and (b), it is evident that for a fixed value of temperature splitting of the two transverse modes increases with increase of chiral chemical potential. Similar conclusion can be made from Figs.~\ref{wtl} (c) and (d). However the increase in temperature leads to decrease in energy gap between two transverse modes as observed by comparing figures in the vertical panels. 

\begin{figure}[h]
	\includegraphics[scale=0.375]{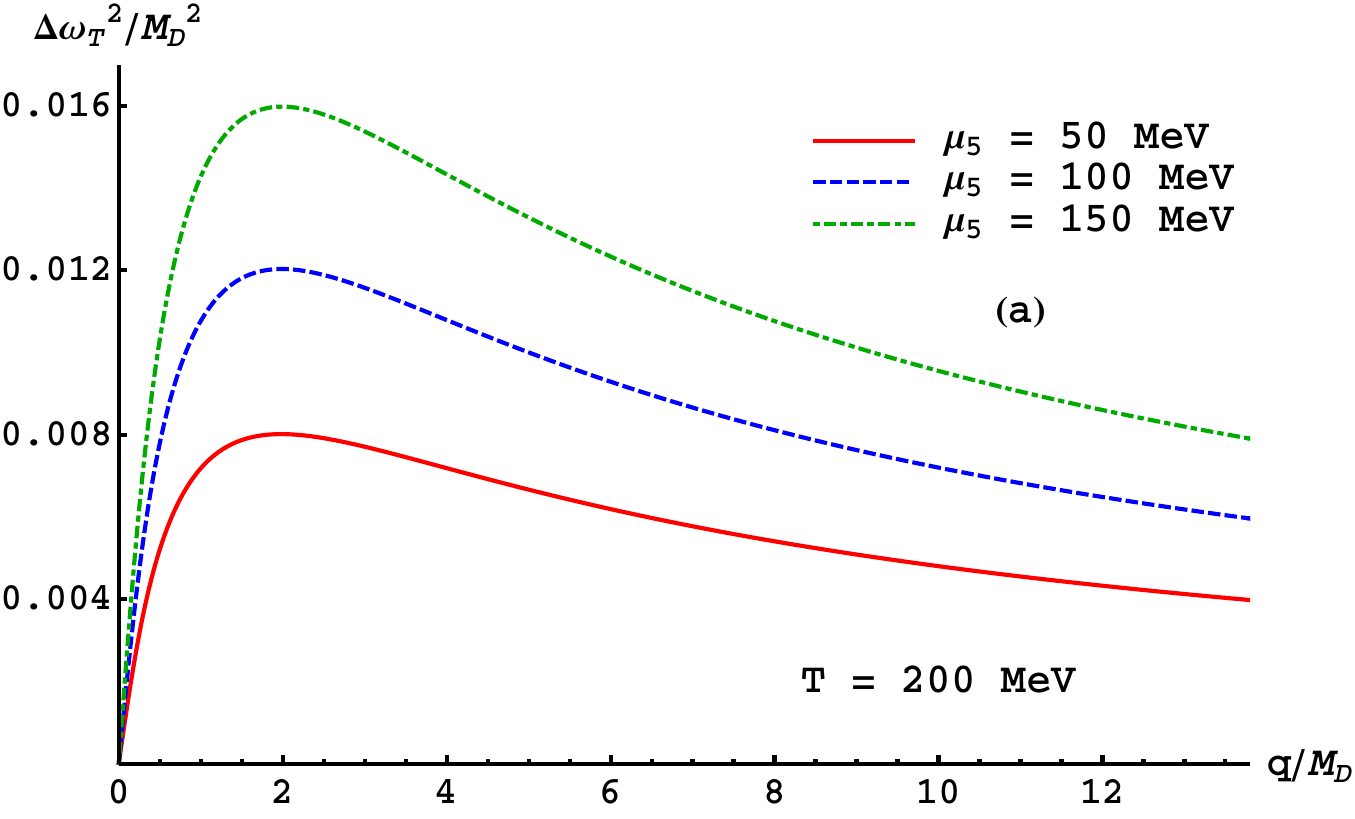} ~~~~
	\includegraphics[scale=0.375]{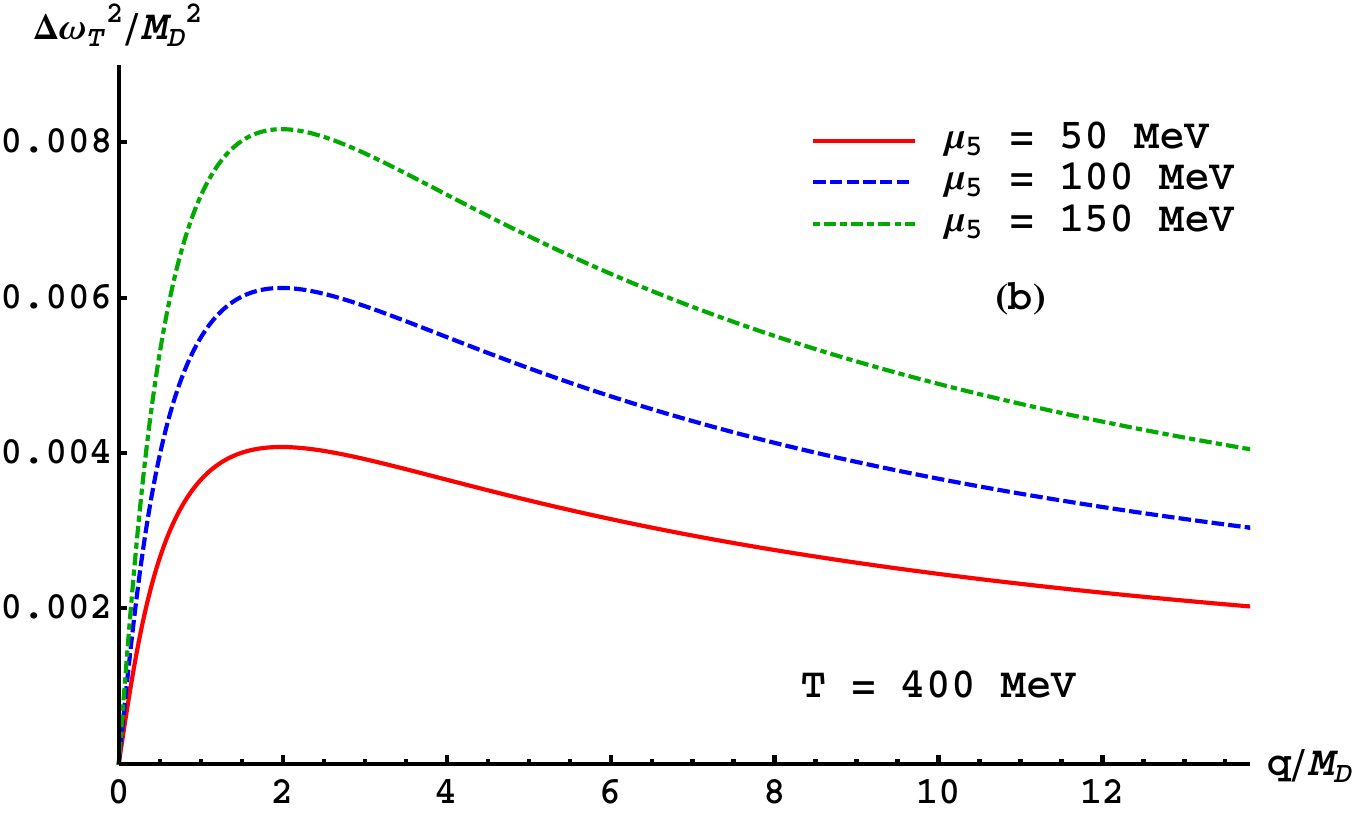}
	\caption{Difference $\Delta\omega_T^2$ as function of the momentum $q$ in units of $\MD$}\label{delomega2}
\end{figure}
To explore the interplay between temperature and CCP in the splitting of two transverse modes in a chirally imbalanced medium, in Figs.~\ref{delomega2}(a) and (b) we have shown the variation of the difference between two squared transverse modes i.e $\Delta\omega^2_T=(\omega_T^+)^2-(\omega_T^-)^2$ normalised by the Debye mass as function of scaled momentum at $ T=200 $ and 400 MeV respectively for different values of $ \mu_5 $. In both the plots this difference starts from zero, attains a maxima and then slowly decreases with increasing momentum. Clearly the amount of splitting is larger (smaller) at smaller (larger) temperature value. 

Now we obtain the approximate analytic solutions of Eqs.~\eqref{dis1} and \eqref{dis2} in the limit of small and large momenta. For small values of momentum $q\ll\MD$
\begin{align}
	\omega_L &\approx \frac{\MD}{\sqrt{3}}\fbr{1+\frac{9}{10}\frac{q^2}{\MD^2}-\frac{27}{280}\frac{q^4}{\MD^4}+\cdots} \label{wl_small}\\
	\omega_T^\pm &\approx \frac{\MD}{\sqrt{3}}\tbr{1\pm\frac{e^2\mu_5 }{4\pi^2 \MD}\frac{q}{\MD}+\fbr{\frac{9}{5}\mp\frac{e^4\mu_5^2 }{32\pi^4 \MD^2}}\frac{q^2}{\MD^2}\pm\fbr{-\frac{9e^2\mu_5 }{10\pi^2 \MD}+\frac{e^6\mu_5^3 }{128\pi^6 \MD^3}}\frac{q^3}{\MD^3}+\cdots}\label{wt_small}~.
\end{align}
Here $ M_D /\sqrt{3}$ is the plasmon mass and classically this is the lowest frequency of
plasma oscillations which is the same for transverse and longitudinal modes. The behaviour of different modes (shown in Figs.~\ref{wtl} (a)$ - $(d)) at large wavelength i.e. small momentum is evident from Eqs.~\eqref{wl_small} and \eqref{wt_small}. For instance it is clear that at $ q\to 0 $, the longitudinal and transverse modes are degenerate. Moreover, it can be seen that for smaller values of momentum all three modes  are relevant.  Similarly for $q\gg\MD$ the approximate analytic solutions for longitudinal and transverse modes are given by
\begin{align}
	\omega_L &\approx  q+2qe^{-2\frac{q^2+\MD^2}{\MD^2}} \label{wl_large}\\
	\omega_T^\pm &\approx  q\tbr{1+\frac{\MD^2}{4q^2}\mp\frac{e^2\mu_5\MD^2}{8\pi^2 q^3}\fbr{1-\frac{1}{2}\ln\frac{8q^2}{\MD^2}}+\cdots}~\label{wt_large}.
\end{align}
The large momentum behaviour of different modes shown in Figs.~\ref{wtl} (a)$ - $(d) can be understood from the above results. For example, for Eq.~\eqref{wl_large} at large values of $ q $ the exponential term will become vanishingly small implying that the longitudinal mode will approach the free dispersion very fast compared to the transverse modes in which different terms are suppressed following a power law. 

\begin{figure}[h]
	\includegraphics[scale=0.375]{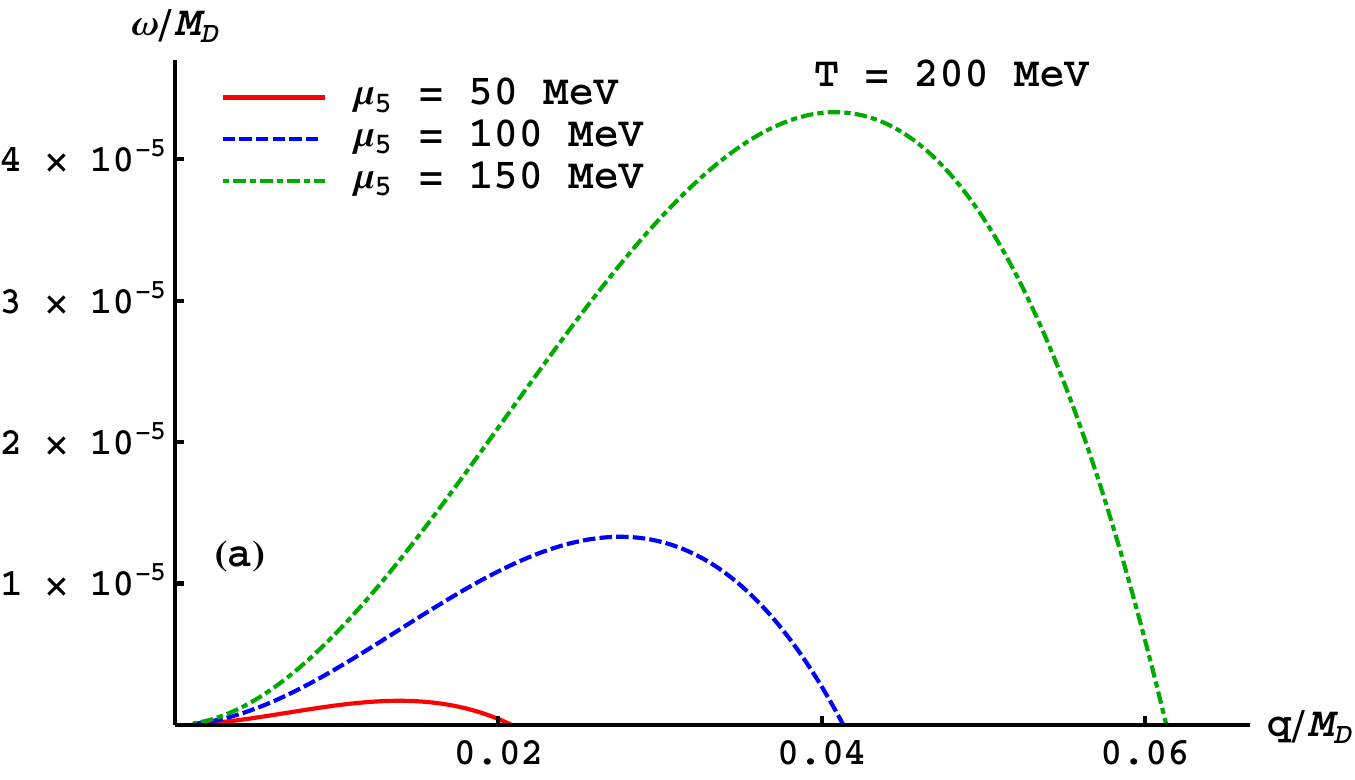} ~~~~
	\includegraphics[scale=0.375]{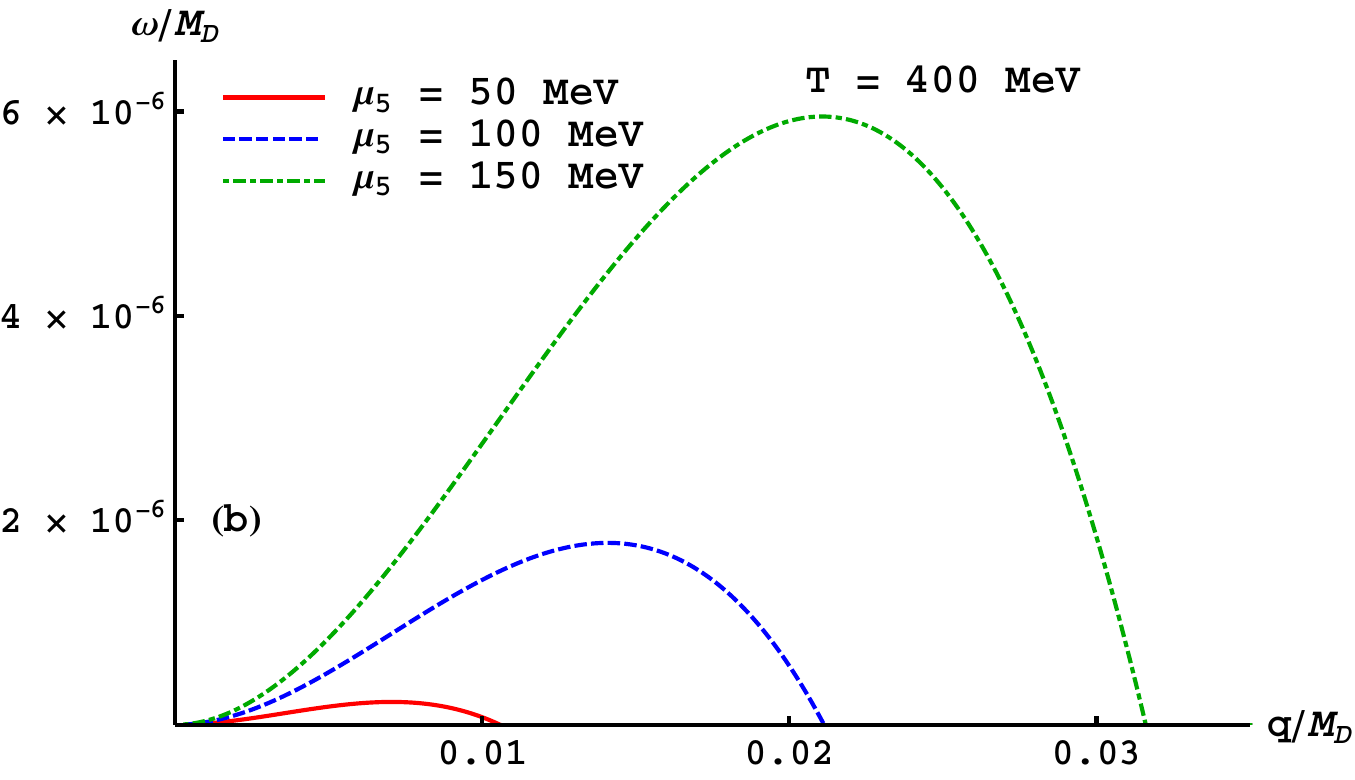}
	\caption{Imaginary parts of the dispersion relation for the unstable mode.} 

	\label{imroot}
\end{figure}

Till now we have discussed the stable modes of the dispersion relations given by Eqs.~\eqref{dis1} and \eqref{dis2} i.e. $ \IM \omega = 0 $ allowing these modes to maintain a constant amplitude. However they are in general complex-valued. If the imaginary part of the frequency of a mode is negative i.e. $ \IM \omega < 0 $, the mode is damped and its amplitude decays exponentially over time as $ \exp \tbr{\fbr{\IM \omega}  t} $. The mode is considered to be over-damped if $ \RE \omega  = 0 $ in addition to $ \IM \omega < 0 $.  On the other hand, if $  \IM \omega > 0 $, the amplitude of the mode grows exponentially in time leading to an instability. In chirally imbalanced medium we find that the dispersion relation given by Eq.~\eqref{dis2} with positive sign has pure imaginary solutions that correspond to unstable modes which grow exponentially and result in instability of the chiral plasma~\cite{Akamatsu:2013pjd,Carignano:2018thu}. 
In the quasistatic limit $\md{\omega}\ll q$ expression for imaginary $\omega$ is given by\cite{Akamatsu:2013pjd}  
\begin{equation}
	\omega=i\gamma~~~~~ \text{with}~~~~~ \gamma=\frac{4\mu_5}{\pi^2 \MD^2}\frac{g^2 N_f}{4\pi}q^2\fbr{1-\frac{4\pi^2 q}{g^2 N_f\mu_5}} \label{Eq_gm}~.
\end{equation}
The value of the imaginary part of $\omega$ is maximum for $q = \frac{g^2 N_f \mu_5}{6\pi^2}$, and its maximum value is given by:
\begin{equation}
	\gamma_{max}=\frac{16\mu_5^3}{27 \pi^4\MD^2}\fbr{\frac{g^2 N_f}{4\pi}}^3\approx 1.314\times10^{-3}\frac{\mu_5^3}{\MD^2} ~ \implies \frac{\gamma_{\text{max}}}{M_D} \ll \frac{\omega_p}{M_D} = \frac{1}{\sqrt{3}}~. \label{Eq_max_gm}
\end{equation}
	Thus one can infer that $\gm_{\rm max}$ is at least 4 orders of magnitude smaller than the plasma frequency of the system. Since the time scale of the plasma instability  $\sim 1/\gm $, it is evident that the instability develops over a much longer time scale compared to the characteristic response time of the system which is the inverse of the plasma frequency~\cite{Akamatsu:2013pjd}. These are plotted in Figs.~\ref{imroot} (a) and (b) for different values of temperature and CCP. It is evident that unstable modes occur at low momenta. Moreover, increase in chiral chemical potential leads to increase in magnitude of $\gm$ at a given temperature. This is understandable for Eq.~\eqref{Eq_gm} as $\gm$ is proportional to $\mu_5$.
	 On the other hand magnitude of $\gm$ decreases in the chiral plasma with increase in temperature as evident from Eq.~\eqref{Eq_max_gm}. This implies that for higher values of $T$ instability develops over a longer timescale. 
	
	\begin{figure}[h]
		\includegraphics[scale=0.41]{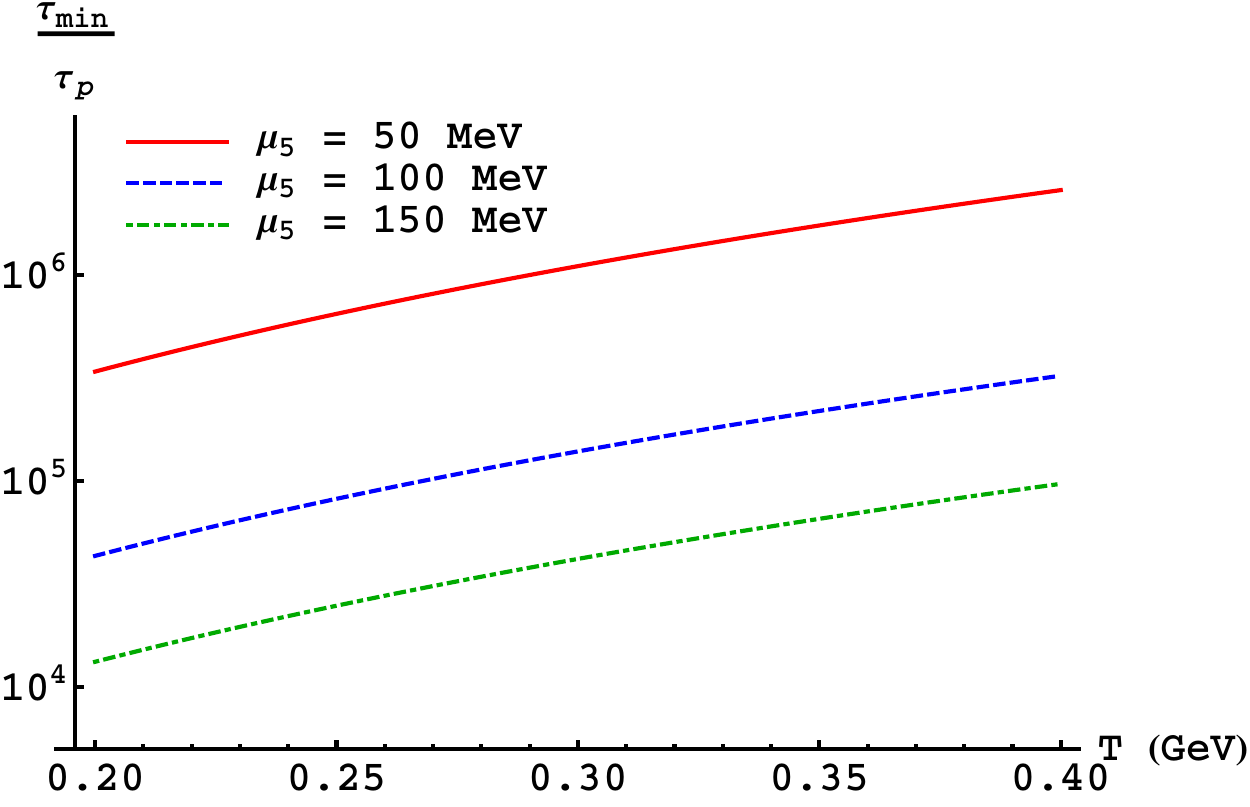}
				\caption{$\tau_{\rm min}/\tau_p$  as a function of temperature for different values of $\mu_5$.} 
		\label{Fig_timescale}
	\end{figure} 
To further investigate the timescale associated with the development of plasma instabilities in Fig.~\ref{Fig_timescale} we have plotted the ratio $\tau_{\rm min}/\tau_p$  as a function of temperature for different values of $\mu_5$. Here $\tau_{\rm min}= 1/\gm_{\rm max}$ corresponds to the minimum time required for the instability to develop whereas $\tau_p $ denotes the characteristic response time of system represented by the inverse of  the plasma frequency. As discussed, it is evident that a longer time is required for the instability to develop for higher values of $\mu_5$. Moreover, it should be noted that for all the cases $\tau_{\rm min}$ exceeds the characteristic response time of the system $\tau_p$ by four to five orders of magnitude. This separation of timescales justifies the definition of the heavy quark static potential in the context of a chiral plasma which will be discussed in the next section.

\section{Complex Heavy Quark Potential}\label{Sec_potential}

	The dissociation of heavy quarkonium has long been proposed as a sensitive probe for studying the hot and dense medium created in high-energy collisions. While bound states of heavy quarks can potentially survive within the quark-gluon plasma even at temperatures above the deconfinement threshold the color screening effect induced by the presence of light quarks and gluons weakens the binding interaction between the quark-antiquark pair. This screening, known as Debye screening ultimately leads to the dissociation of quarkonium states. 	
	The static heavy-quark (HQ) potential due to one gluon exchange in the real-time formalism can be determined through the Fourier transform of the 11-component of the effective gluon propagator~\cite{Laine:2006ns,Dumitru:2009fy,Guo:2018vwy,Dong:2022mbo}
	\begin{equation}    
		V(r)= 4\pi\alpha_s\int \frac{d^3q}{(2\pi)^3}\fbr{e^{i\vec{q}\cdot\vec{r}}-1}G^{00}(q^0\rightarrow 0,\vec{q})_{11}~.
	\end{equation}
	Real part of this potential gives the Debye screened potential and imaginary part determines the decay width of a quarkonium state.
	
	\begin{figure}[h]
		\includegraphics[scale=0.41]{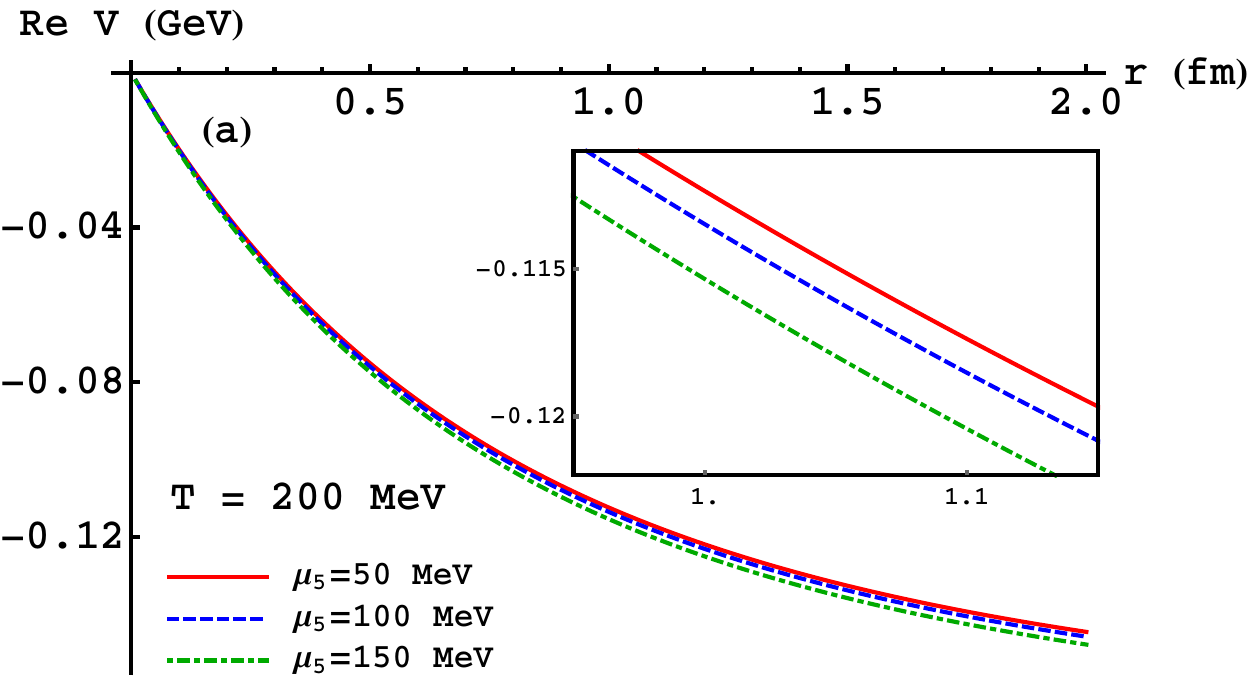} ~~~~~
		\includegraphics[scale=0.41]{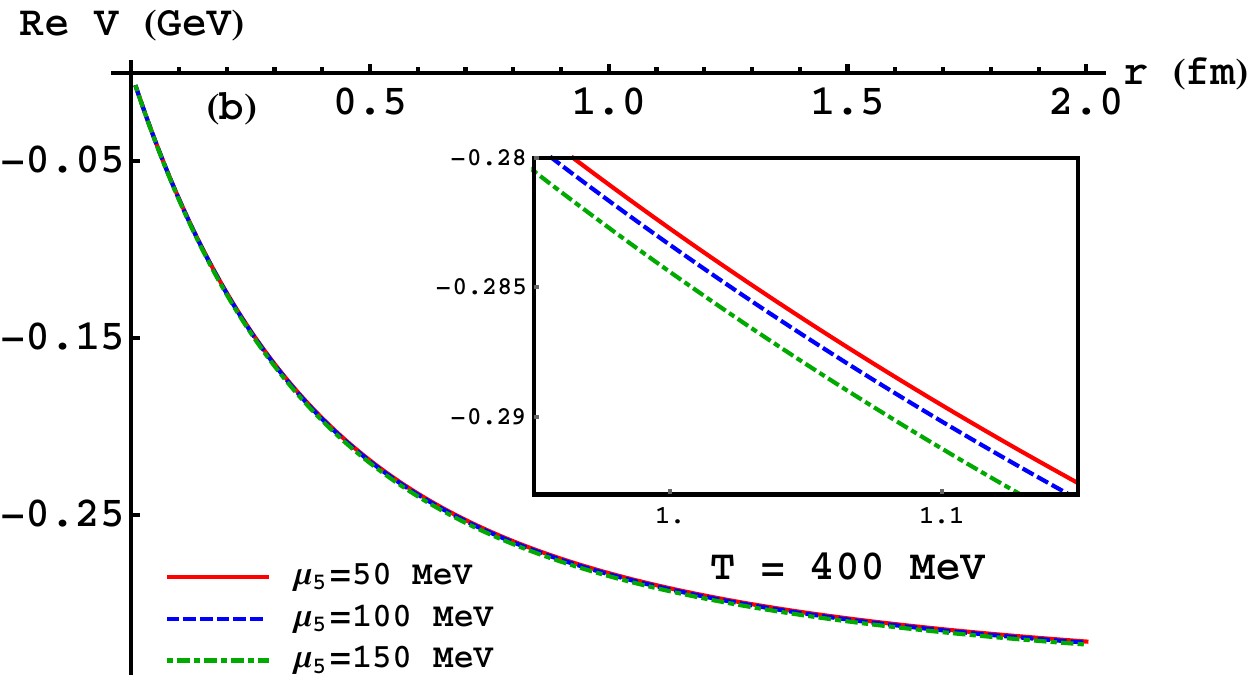}
		\caption{The real part of the heavy-quark potential as a function of r, evaluated at three different chemical potentials ($\mu_5$ = 50, 100, and 150 MeV) for $T = 200$ MeV (left plot, Fig. a) and $T = 400$ MeV (right plot, Fig. b). For this graph, we take $\alpha = 0.4$.} 
		\label{ReV}
	\end{figure}	
	Now the real part of $G^{00}_{11}$ can be calculated starting from Eq.~\eqref{Eff_prop} using Eq.~\eqref{Eq_Re11_Re}. It follows that 
	\begin{align}
		\text{Re }G^{00}(q^0\rightarrow 0,\vec{q})_{11}&=\text{Re }G^{00}(q^0\rightarrow 0,\vec{q})= \frac{-1}{q^2+\MD^2}
	\end{align}	
	So real part of the potential is given by
	\begin{equation}
		\text{Re }V(r)= -4\pi\alpha_s\int \frac{d^3q}{(2\pi)^3}\fbr{e^{i\vec{q}\cdot\vec{r}}-1}\fbr{\frac{1}{q^2+\MD^2}-\frac{1}{q^2}}=\alpha_s \fbr{\frac{1-e^{-\frac{r}{r_D}}}{r}-\frac{1}{r_D}}
	\end{equation} 
	where $r_D=\MD^{-1}$ is called Debye radius. For a fixed temperature and chemical potential, the screening mass (or Debye mass) increases with increasing chiral chemical potential. As a result, screening occurs over shorter distances for higher chiral chemical potentials, which is understood by plotting the real part of the potential as a function of distance, as shown in Fig.~\ref{ReV}.
	
	\begin{figure}[h]
		\includegraphics[scale=0.41,angle=0]{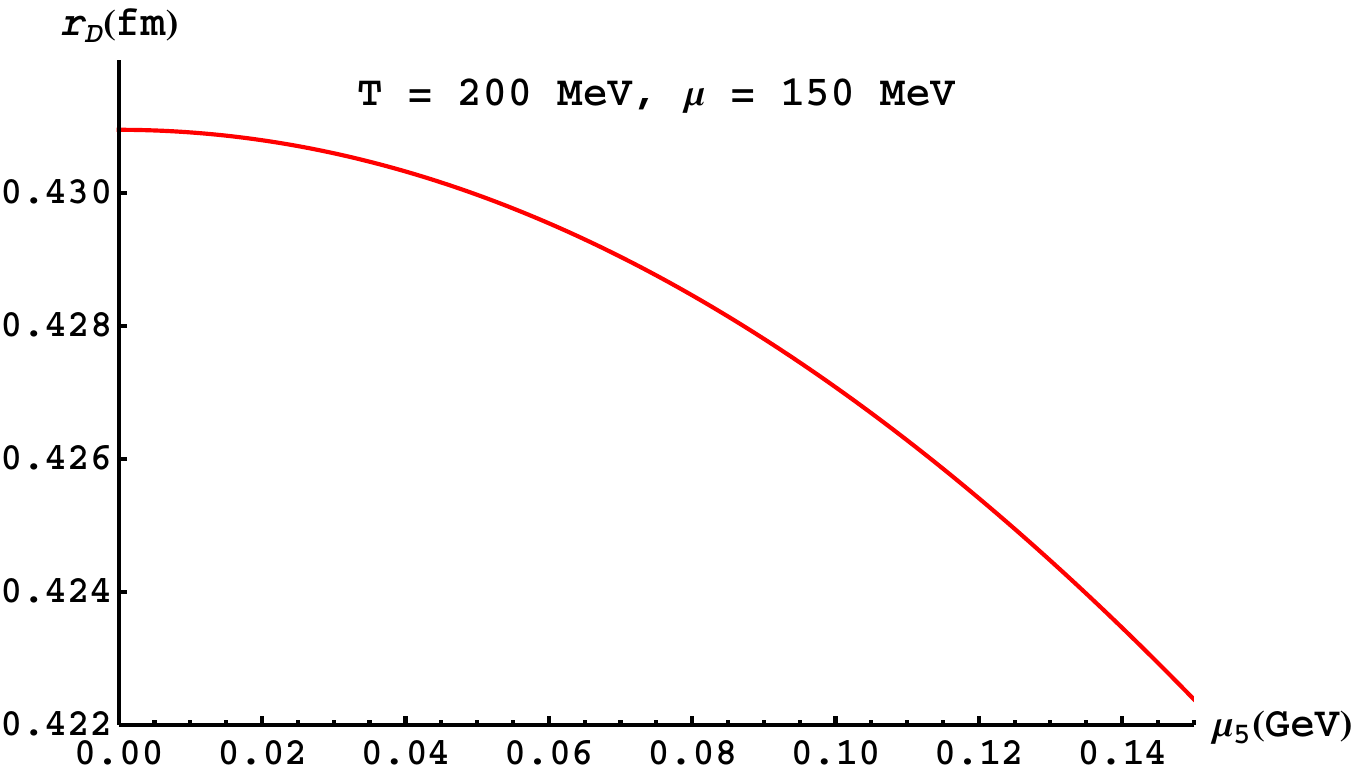} 
		\caption{The variation of Debye radius as a function of chiral chemical potential.\label{Fig_rd}}
		
	\end{figure}
	To further quantify this effect, the Debye radius as a function of chiral chemical potential at $T=200$ MeV and $\mu=150$ MeV is shown in  Fig.~\ref{Fig_rd}. It is seen that the Debye radius decreases with increase in chiral chemical potential which restricts the formation of heavy quarkonia in a chirally asymmetric medium.

\begin{figure}[h]
	\includegraphics[scale=0.41]{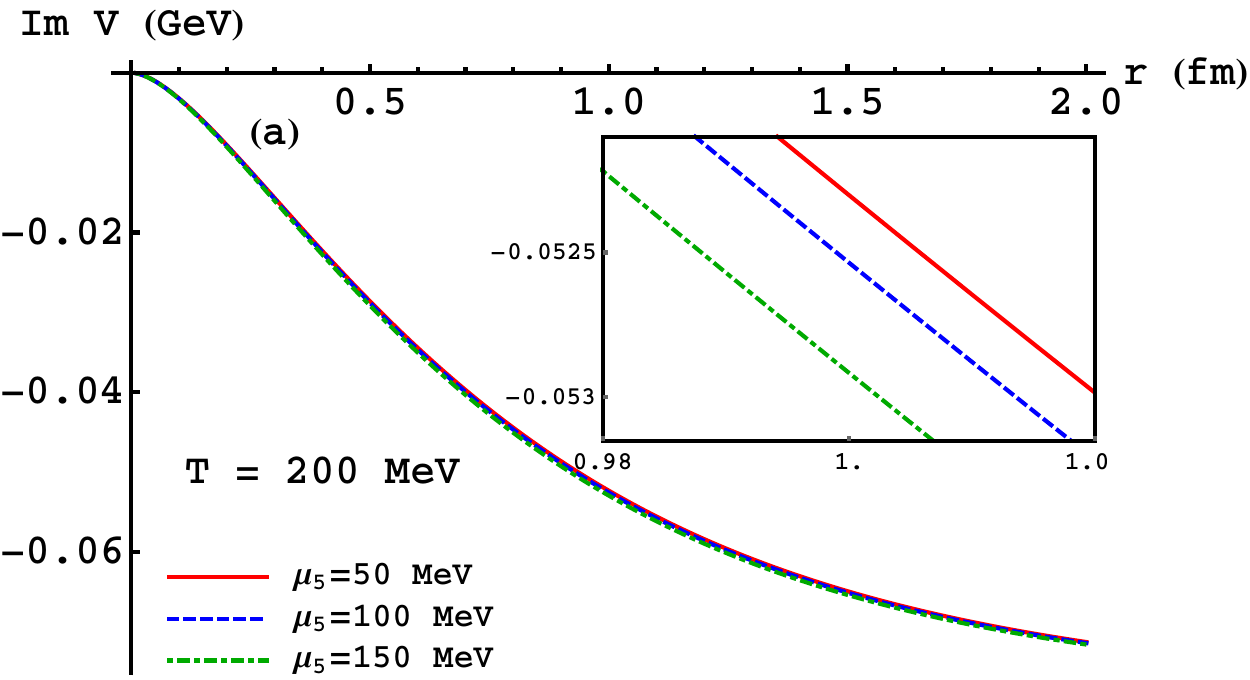} ~~~~~
	\includegraphics[scale=0.41]{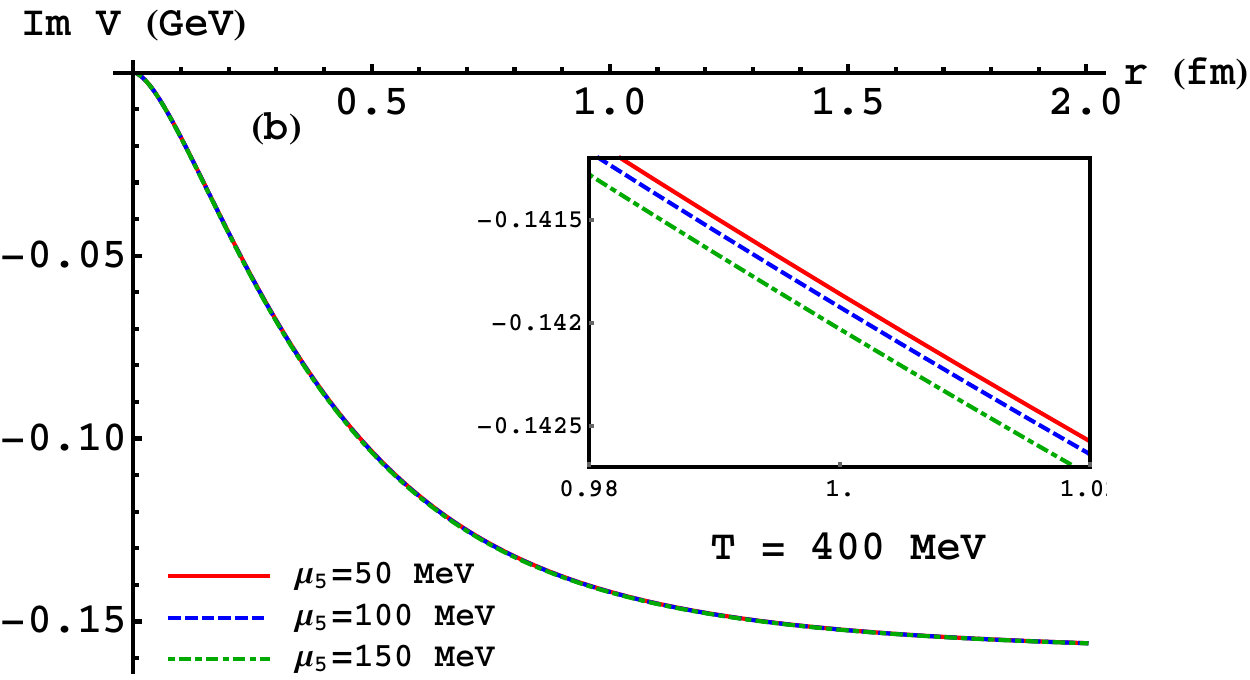}
	\caption{The imaginary part of the heavy-quark potential as a function of r, computed at three values of the chiral chemical potential ($\mu_5$ = 50, 100, and 150 MeV), is shown for $T = 200$ MeV (left plot, Fig. a) and $T = 400$ MeV (right plot, Fig. b). The parameter $\alpha$ is set to 0.4 in this analysis.} 
	
	\label{ImV}
\end{figure}

	The imaginary part of $G^{00}(q_0, \vec{q})_{11}$ can be obtained from Eq.~\eqref{Eff_prop} by using Eq.~\eqref{Eq_Im11_Im} and we get
	\begin{align}
	\text{Im }G^{00}(q^0\rightarrow 0,\vec{q})_{11}&=\lim_{q^0\rightarrow 0}\coth(\frac{\beta q^0}{2})\text{Im }G^{00}(q^0,q)=\frac{\pi T \MD^2}{q(q^2+\MD^2)^2}~,
\end{align}	
where we have used the relation
\begin{equation}
	\log\frac{q^0+q+i\epsilon}{q^0-q+i\epsilon}=\log\md{\frac{q^0+q}{q^0-q}}-i\pi\Theta(q^2-(q^0)^2)~.
\end{equation}
Thus the imaginary part of the potential is expressed as
\begin{equation}
	\text{Im }V(r)= 4\pi\alpha_s\int \frac{d^3q}{(2\pi)^3}\fbr{e^{i\vec{q}\cdot\vec{r}}-1}\fbr{\frac{\pi T \MD^2}{q(q^2+\MD^2)^2}}=\alpha_s T \fbr{\phi_2\fbr{\frac{r}{r_D}}-1}
\end{equation}
where 
\begin{equation}
	\phi_2\fbr{\frac{r}{r_D}}=4\pi^2\int \frac{d^3(q/ \MD)}{(2 \pi)^3}e^{i\vec{q}\cdot\vec{r}}\frac{1}{\frac{q}{\MD}\sbr{\fbr{\frac{q}{\MD}}^2+1}^2}
\end{equation}
with
\begin{equation}
	\phi_n(r)=2\int_{0}^{\infty}dz\frac{\sin(zr)}{zr}\frac{z}{(z^2+1)^n}~.
\end{equation}
For fixed values of temperature and chemical potential, the magnitude of the imaginary part of the heavy quark potential grows with increasing chiral chemical potential, as shown in Fig.~\ref{ImV}. This behaviour implies that heavy quarkonium states experience faster decay rates in a medium with higher chiral chemical potential. Note that, all the analytical  expressions provided in this section are in agreement with~\cite{Dumitru:2009fy,Guo:2018vwy,Dong:2022mbo} for vanishing $\mu_5$ in absence of spatial asymmetry.

\section{sum rules and residues}	

The longitudinal and transverse gluon propagators $\Delta_L$ and $\Delta_T^\pm$ are defined as
\begin{eqnarray}
	\Delta_L(q_0,q)&&=\frac{Q^2}{q^2}\frac{-1}{Q^2+\Pi_L}\nn\\
	&&=\frac{-1}{q^2+\MD^2\fbr{1-\frac{q_0}{2q}\ln\frac{q_0+q}{q_0-q}}}
\end{eqnarray}
and
\begin{eqnarray}
	\Delta_T^\pm(q_0,q)&&=\frac{-1}{Q^2+\Pi_T\mp\Pi_A}\nn\\
	&&=\frac{-1}{q_0^2-q^2-\frac{1}{2}\MD^2 \frac{q_0^2}{q^2}\fbr{1+\frac{1}{2}\fbr{\frac{q}{q_0}-\frac{q_0}{q}}\logterm}\pm\frac{g^2}{2\pi^2}\frac{N_f}{2} \mu_5 \frac{q_0^2-q^2}{q}\fbr{1-\frac{q_0}{2q}\logterm}}~.	
\end{eqnarray}
	
\begin{figure}[h]
	\includegraphics[scale=0.34,angle=0]{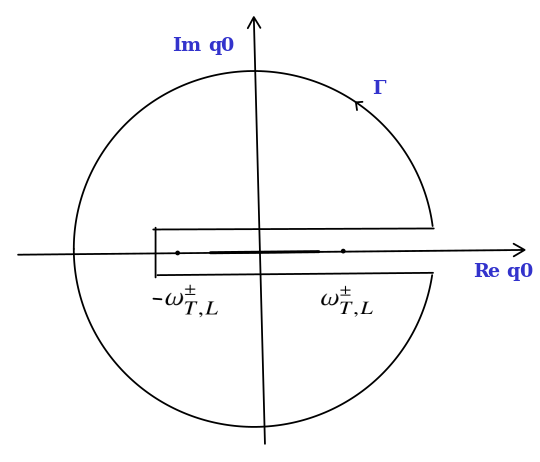} 
	\caption{Integration contour for derivation of sum rules}\label{contour}
	\end{figure}
$\Delta_L$ and $\Delta_T^\pm$ are analytic in the complex $q_0$-plane cut from $-q$ to $q$ and they have poles at $q_0=\pm\omega_L$ and $\pm\omega_T^\pm$. So one can write for $\Delta=\Delta_L$ or $\Delta_T^\pm$ using Cauchy's theorem 
\begin{eqnarray}\label{cauchy}
	\Delta(q_0,q)&&=\oint_\Gamma \frac{dz}{2\pi i}\frac{\Delta(z,q)}{z-q_0}\nn\\
	&&=\int_{-\infty}^{\infty}\frac{dq_0^\prime}{2\pi i}\frac{\Delta(q_0^\prime+i\ep,q)-\Delta(q_0^\prime-i\ep,q)}{q_0^\prime-q_0}+\oint_{\Gamma^\prime} \frac{dz}{2\pi i}\frac{\Delta(z,q)}{z}~,
\end{eqnarray}
where $\Gamma$ is the contour shown in Fig.~\ref{contour} and $\Gamma^\prime$ is a circle whose radius tends to infinity. Eq.~\ref{cauchy} can be written in terms of spectral density $\rho(q_0,q)=2\text{Im}\Delta(q_0+i\ep,q)$ which contains both the discontinuities across the cuts and residues at poles~\cite{Bellac:2011kqa}. The  above equation can be written as
\begin{equation}\label{delq}
	\Delta(q_0,q)=\int_{-\infty}^{\infty}\frac{dq_0^\prime}{2\pi }\frac{\rho(q_0^\prime,q)}{q_0^\prime-q_0}+\oint_{\Gamma^\prime} \frac{dz}{2\pi i}\frac{\Delta(z,q)}{z}	
\end{equation}	
Now $\Delta_T^\pm(z,q)\sim\frac{1}{z^2}$. As $z\rightarrow\infty$, $\Delta_T^\pm(z,q)\rightarrow0$, there is no contribution from $\Gamma^\prime$. For $z\rightarrow\infty$, $\Delta_L(z,q)\rightarrow\frac{-1}{q^2}$ and this will contribute for contour $\Gamma^\prime$. Sum rules are obtained by setting $q_0=0$ in the Eq.~\ref{delq}~\cite{Bellac:2011kqa}.
Sum rule for $\Delta_L$ at small $q_0$ is given by
\begin{equation}
	\int_{-\infty}^{\infty}\frac{dq_0}{2\pi }\frac{\rho_L(q_0,q)}{q_0-q_0}=\frac{1}{q^2}+\Delta_L(0,q)=\frac{\MD^2}{q^2\fbr{q^2+\MD^2}}
\end{equation}
and the sum rule for $\Delta_T^\pm$ at small $q_0$ is
\begin{equation}
	\int_{-\infty}^{\infty}\frac{dq_0}{2\pi }\frac{\rho_T^\pm(q_0,q)}{q_0-q_0}=\Delta_T^\pm(0,q)=\frac{1}{q^2\pm\frac{g^2}{2\pi^2}\frac{N_f}{2} \mu_5}~.
\end{equation}

Other sum rules may be obtained by setting $q_0$ very large~\cite{Bellac:2011kqa}. This is given by
\begin{eqnarray}\label{delL}
	\Delta_L(q_0,q)&&=-\frac{1}{q^2}+\int_{-\infty}^{\infty}\frac{dq_0}{2\pi }\frac{\rho_L(q_0,q)}{q_0-q_0}\nn \\
	&&=-\frac{1}{q^2}-\frac{1}{q_0}\sum_{n=0}^{\infty}\int_{-\infty}^{\infty}\frac{dq_0^\prime}{2\pi}\fbr{\frac{q_0^\prime}{q_0}}^{2n+1}\rho_L(q_0^\prime,q)~.
\end{eqnarray}
Now expanding $\Delta_L(q_0,q)$ for large $q_0$ we obtain
\begin{equation}\label{delL1}
	\Delta_L(q_0,q)=-\frac{1}{q^2}-\frac{\MD^2}{3q^2q_0^2}-\frac{\frac{\MD^2}{5}+\frac{\MD^4}{9q^2}}{q_0^4}+\mathcal{O}[q_0^{-6}]~.
\end{equation}
Comparing the powers of $q_0^{-2}$ on the right hand sides of the Eqs.~\eqref{delL} and \eqref{delL1} we get
\begin{equation}
	\int_{-\infty}^{\infty}\frac{dq_0}{2\pi}q_0\rho_L(q_0,q)=\frac{\MD^2}{3q^2}=\frac{\omega_p^2}{q^2}~.
\end{equation}
Similarly comparison of the power of $q_0^{-4}$ on the right hand sides of the Eqs.~\eqref{delL} and \eqref{delL1} gives
\begin{equation}
	\int_{-\infty}^{\infty}\frac{dq_0}{2\pi}q_0^3\rho_L(q_0,q)=\frac{\MD^2}{5}+\frac{\MD^4}{9q^2}=\frac{3\omega_p^2}{5}+\frac{\omega_p^4}{q^2}~.
\end{equation}

Now we proceed to calculate the sum rules for $\Delta_T^\pm$ at large $q_0$ where
\begin{eqnarray}\label{delT}
	\Delta_T^\pm(q_0,q)&&=\int_{-\infty}^{\infty}\frac{dq_0}{2\pi }\frac{\rho_T^\pm(q_0,q)}{q_0-q_0}\nn \\
	&&=-\frac{1}{q_0}\sum_{n=0}^{\infty}\int_{-\infty}^{\infty}\frac{dq_0^\prime}{2\pi}\fbr{\frac{q_0^\prime}{q_0}}^{2n+1}\rho_T^\pm(q_0^\prime,q)
\end{eqnarray}
Now expanding $\Delta_T^\pm(q_0,q)$ for large $q_0$ we get
\begin{equation}\label{delT1}
	\Delta_T^\pm(q_0,q)=-\frac{1}{q_0^2}-\frac{1}{q_0^4}\tbr{\frac{\MD^2}{3}+q^2\pm\frac{g^2}{12\pi^2}\frac{N_f}{2}\mu_5q}+\mathcal{O}[q_0^{-6}]~.
\end{equation}

Comparing the powers of $q_0^{-2}$ on the right hand sides of the Eqs.~\eqref{delT} and \eqref{delT1} we get
\begin{equation}
	\int_{-\infty}^{\infty}\frac{dq_0}{2\pi}q_0\rho_T^\pm(q_0,q)=1
\end{equation}
and comparison of the power of $q_0^{-4}$ on the right hand sides of the Eqs.~\eqref{delT} and \eqref{delT1} gives
\begin{equation}
	\int_{-\infty}^{\infty}\frac{dq_0}{2\pi}q_0^3\rho_T^\pm(q_0,q)=\frac{\MD^2}{3}+q^2\pm\frac{g^2}{12\pi^2}\frac{N_f}{2}\mu_5q=\omega_p^2+q^2\pm\frac{g^2}{12\pi^2}\frac{N_f}{2}\mu_5q~.
\end{equation}

Now residues of $\Delta_L$ and $\Delta_L^\pm$ are evaluated at the poles.
Residue of the longitudinal excitations is given by
\begin{equation}
	Z_L(q)=-\fbr{\frac{\partial\Delta_L^{-1}(q_0,q)}{\partial q_0}\Big{\arrowvert}_{q_0=\omega_L(q)}}^{-1}=\frac{\omega_L\fbr{\omega_L^2-q^2}}{q^2\fbr{q^2+\MD^2-\omega_L^2}}
\end{equation}
and residues of the corresponding expressions for transverse modes are given by
\begin{equation}
	Z_T^\pm(q)=-\fbr{\frac{\partial(\Delta_T^\pm)^{-1}(q_0,q)}{\partial q_0}\Big{\arrowvert}_{q_0=\omega_T^\pm(q)}}^{-1}=\frac{\omega_T^\pm\fbr{(\omega_T^\pm)^2-q^2}}{(\omega_T^\pm)^2\MD^2-\tbr{(\omega_T^\pm)^2-q^2}^2\pm\frac{g^2}{2\pi^2}\frac{N_f}{2} \mu_5 q \tbr{(\omega_T^\pm)^2-q^2}}~.
\end{equation}
So the complete expression of the spectral function for longitudinal excitations is given by
\begin{equation}
	(2\pi)^{-1}\rho_L(q_0,q)=Z_L(q)[\delta(q_0-\omega_L(q))-\delta(q_0+\omega_L(q))]+\beta_L(q_0,q)
\end{equation}
where $\beta_L(q_0,q)$ is
\begin{equation}
	\beta_L(q_0,q)=\frac{\frac{1}{2}\MD^2\frac{q_0}{q}\Theta\fbr{1-\frac{q_0}{q}}}{\tbr{q^2+\MD^2\fbr{1-\frac{q_0}{2q}\logterm}}^2+\frac{\MD^4\pi^2q_0^2}{4q^2}}
\end{equation}
and the complete expression of the spectral function of transverse excitations is given by
\begin{equation}
	(2\pi)^{-1}\rho_T^\pm(q_0,q)=Z_T^\pm(q)[\delta(q_0-\omega_T^\pm(q))-\delta(q_0+\omega_T^\pm(q))]+\beta_T^\pm(q_0,q)
\end{equation}
where $\beta_T^\pm(q_0,q)$ is given by
\begin{equation}
	\beta_T^\pm(q_0,q)=\frac{\frac{1}{2}\Theta\fbr{1-\frac{q_0^2}{q^2}}\frac{q_0q}{q_0^2-q^2}\fbr{-\frac{\MD^2}{2}\pm \frac{g^2}{2\pi^2}\frac{N_f}{2} \mu_5 q }}{\tbr{q^2-\frac{1}{2}\frac{\MD^2q^2}{q_0^2-q^2}+\fbr{-\frac{\MD^2}{2}\pm \frac{g^2}{2\pi^2}\frac{N_f}{2} \mu_5 q }\fbr{1-\frac{q_0}{2q}\logterm}}^2+\frac{\pi^2q_0^2}{4q^2}\fbr{-\frac{\MD^2}{2}\pm \frac{g^2}{2\pi^2}\frac{N_f}{2} \mu_5 q }^2}~.
\end{equation}	


\section{Summary and conclusion}

Gluon self energy is studied in a chirally imbalanced plasma using hard thermal loop approach in the real time formulation of thermal field theory.  
Real poles of the effective propagator give stable modes. These stable modes split into three modes, one is longitudinal and others are transverse modes. Again, the transverse modes split into left and right handed circularly polarized modes due to the chiral asymmetry. 
Imaginary poles of the propagator give unstable modes which are responsible for instability in the plasma which occur at very low momenta with respect to Debye mass. Examination of the timescale associated with development of the instabilities reveals that the minimum time required for the instability to develop always  exceeds the characteristic response time of the system i.e. the inverse of the plasma frequency. 
Moreover it was observed that with increasing temperature (chiral chemical potential) the magnitude of the unstable mode  decreases (increases) in the chiral plasma. This, in turn, implies that  the instability develops over a longer (shorter) timescale at higher temperatures (chiral chemical potential).
Both real and imaginary parts of the complex static heavy quark potential were investigated. It is seen that for a fixed temperature and chemical potential, the screening mass (or Debye mass) increases with increasing chiral chemical potential. As a consequence, screening occurs over shorter distances for higher chiral chemical potentials. 
The characteristic length scale of this chirally asymmetric plasma i.e Debye radius decreases with the chiral chemical potential inhibiting the formation of heavy quarkonia.
 Furthermore, the magnitude of the imaginary part of the heavy quark potential grows with increasing chiral chemical potential for fixed values of temperature and chemical potential implying faster decay of heavy quarkonium states in a medium with higher chiral asymmetry. 
 Spectral density of gluon is also evaluated. Properties of spectral function such as sum rules and residues for several cases are obtained which will be necessary to calculate energy loss, damping rate and transport coefficients such as conductivity, viscosity etc.

\appendix

\section{The real time formulation of thermal field theory}
\label{App_RTF}

\begin{figure}[h]
	\begin{center}
		\includegraphics[scale=0.4]{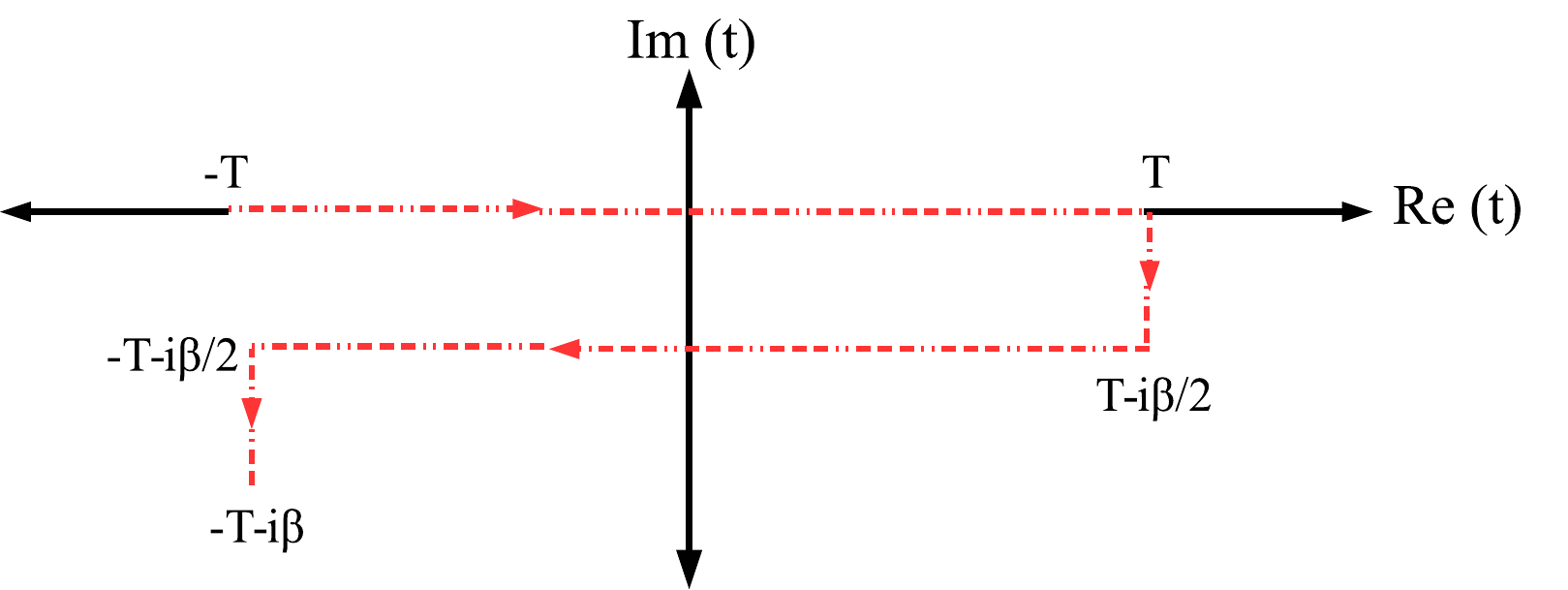}
	\end{center}
	\caption[The contour in the complex time plane for Real Time Formalism]
	{The contour $C$ in the complex time plane for RTF with $T\rightarrow\infty$}
	\label{fig.2.timecontour}
\end{figure}

In this appendix we briefly summarize few important features in the real time formulation of thermal field theory we are using in this article. The free thermal propagator for a general field $\psi_l\FB{X}$ may be defined as
\begin{equation}
	D_{ll'}\FB{X,X'} = i\ensembleaverage{\mcTc\left\{ \psi_l\FB{X}\psi^\dagger_{l'}\FB{X'}  \right\}}
\end{equation}
where $\ensembleaverage{}$ denotes ensemble average. The temporal component of $X$ and $X'$ denoted by $\tau$ and 
$\tau'$ respectively take complex values on the 
contour $C$ in complex time plane. The ensemble average of products of various fields converges in the domain 
$-\beta \le \IM\FB{\tau-\tau'} \le 0$ where $\beta$ is the inverse temperature. This condition defines the region of 
analyticity and puts restriction on the time contour $C$. We will work with the time contour 
as shown in Fig.~\ref{fig.2.timecontour}.
This particular choice of the time contour breaks up the propagator into pieces with points on different segment of 
the contour. If $\tau_h$ and $\tau_v$ are points respectively on the horizontal and vertical segments, the pieces, 
$D_{ll'}\FB{\vec{x},\vec{x'};\tau_h,\tau_v} \rightarrow 0$ by Riemann-Lebesgue lemma. In the context of perturbative 
calculations, the contour essentially reduces to two parallel lines, one being the real axis and the other shifted from it 
by $-i\beta/2$. The points on these two lines are denoted by subscripts $1$ and $2$ respectively so that $\tau_1=t$ and 
$\tau_2=t-i\beta/2$. The propagator then consists of four pieces and may be put into the form of $2\times 2$ matrix 
and each of them can be Fourier transformed as
\begin{equation}\label{Eq_Dllp}
	\TB{ \begin{array}{cc}
			D_{ll'}\FB{\vec{x},\vec{x'};t,t'} & D_{ll'}\FB{\vec{x},\vec{x'};t,t'-i\beta/2} \\
			D_{ll'}\FB{\vec{x},\vec{x'};t-i\beta/2,t'} & D_{ll'}\FB{\vec{x},\vec{x'};t-i\beta/2,t'-i\beta/2}
	\end{array} }  = \int\frac{d^4 K}{\FB{2\pi}^4}e^{-i K\cdot (X-X')}\bm{D}_{ll'}(k_0,\vec{k})
\end{equation}
where $\bm{D}_{ll'}\FB{k_0,\vec{k}}$ is the momentum space thermal propagator matrix. From now on $2 \times 2$ matrices in this Appendix  will be denoted by bold face letters. The free thermal propagator has the following factorized form
\begin{equation}
	\bm{ D}^{(0) }_\llp= \bm{U} (k_0)	\TB{\begin{array}{cc}
			{\overline{D}}^{(0)}_\llp & 0\\
			0 & - {\overline{D}^{(0)}_\llp}^* 
	\end{array} } \bm{U} (k_0)
\end{equation}
where $-i \overline{D}^{(0)}_\llp$ is the corresponding vacuum propagator and the matrix $\bm{U} (k_0)$ is built up from the distribution functions. Such factorized form of propagator remains the same for different spins however the diagonalizing matrix changes and contains  the chemical potential for conserved charge. The exact form of different momentum space thermal propagators for different spins are given in~\cite{Mallik:2016anp}.

Now from its spectral representation, the complete propagator $\bm{\mathfrak{D}} $ can also be shown to admit a similar factorisation~\cite{Kobes:1985kc,Bellac:2011kqa,Mallik:2016anp}
\begin{equation}\label{Eq-complete_propagator}
	\bm{\mathfrak{D}}(k_0,\vec{k})_\llp = \bm{U} (k_0)	\TB{\begin{array}{cc}
			\overline{\mathfrak{D}}_\llp & 0\\
			0 & - {\overline{\mathfrak{D}}_\llp}^* 
	\end{array} } \bm{U} (k_0) .
\end{equation}
The complete propagator $\bm{\mathfrak{D}}\FB{k_0, \vec{k}}$ is obtained perturbatively by summing up loop diagrams leading to the Dyson-Schwinger equation
\begin{equation}\label{Eq_Dyson}
	-i\bm{\mathfrak{D}}_\llp\FB{k_0, \vec{k}} =-i \bm{D}^{(0)}_\llp\FB{k_0,\vec{k}} + \SB{ -i \bm{D}^{(0)}_{l k}\FB{k_0,\vec{k}} } \SB{ -i\bm{\Pi}_{k k'}\FB{k_0, \vec{k}} } \SB{ -i\bm{\mathfrak{D}}_{k' l'}\FB{k_0, \vec{k}}}.
\end{equation}
Here $\bm{\Pi}_{k k'}$ is the 1-loop thermal self energy matrix. Now comparing Eqs.~\eqref{Eq-complete_propagator} and \eqref{Eq_Dyson} it follows that $\bm{U}^{-1}(k_0) \bm{\Pi}^{k k'} \bm{U}^{-1}(k_0) $ must have the diagonal structure
\begin{equation}
	\bm{U}^{-1}(k_0) \bm{\Pi}_{k k'} \bm{U}^{-1}(k_0) =\TB{\begin{array}{cc}
			\overline{\Pi}_{k k'} & 0\\
			0 & - {\overline{\Pi}_{k k'}}^* 
	\end{array} } .
\end{equation}
This allows one to write the matrix equation \eqref{Eq-complete_propagator} as an ordinary equation
\begin{eqnarray}
	&& \overline{\mathfrak{D}}_\llp = \overline{D}^{(0)}_\llp - \overline{D}^{(0)}_{l k} ~\overline{\Pi}_{k k'} ~\overline{\mathfrak{D}}_{k' l'} \nn \\
	&\implies& 	\overline{\mathfrak{D}}^{-1}_\llp = [\overline{D}^{(0)}]^{-1}_\llp + \overline{\Pi}_{l l'} 
\end{eqnarray}
and its complex conjugate. We can get the function $ \overline{\Pi}^{k k'}$ entirely from any component of $\bm{\Pi}^{k k'}$, say $11$-component, via the following relations~\cite{Mallik:2016anp}
\begin{eqnarray}
	\text{Re} \overline{\Pi}_{k k'} &=& \text{Re} \fbr{\Pi_{k k'}}_{11} \label{Eq_Re11_Re}\\
	\text{Im} \overline{\Pi}_{k k'} &=& \begin{cases}
		\tanh \FB{\beta k^0/2}\text{Im} \fbr{\Pi_{k k'}}_{11} ~~~~~~~~\text{bosonic} \vspace{1 em} \\
		
		\coth \FB{\beta k^0/2}\text{Im} \fbr{\Pi_{k k'}}_{11} ~~~~~~~~\text{fermionic} \label{Eq_Im11_Im}
	\end{cases}
\end{eqnarray}

\section{Retarded correlation function}
This appendix discusses the relation of the  retarded correlation function to the time ordered one i.e. the 11-component we have been using through out this article. Here we will consider two local operators $\Ja$ and $\Jbd$ (either composite  or fundamental) which are bosonic in nature. Similar calculation can be done for case of fermions as well~\cite{Laine:2016hma}. In the following we will define different correlation functions.

The complex time-ordered correlation function is given by
\begin{equation}
	\Dab (K) = i \Xint \FT{K}{X} \ensembleaverage{\mathcal{T}_c \Ja (X) \Jbd (0)} \label{Eq_2PC}
\end{equation}
where $\mathcal{T}_c$ refers to the time ordering according to Schwinger-Keldysh contour as shown in Fig.~\ref{fig.2.timecontour}. As discussed in Appendix~\Ref{App_RTF}, $\Dab$ will be a $2 \times 2$  matrix as given by Eq.~\eqref{Eq_Dllp}. Now we define various correlation functions as
\begin{align}
	\DR (K) &= i \Xint \FT{K}{X} \theta(t) \ensembleaverage{\TB{\Ja(x), \Jbd (0)}}  \label{Eq_Ret}\\
	\DG (K) &=  \Xint \FT{K}{X}  \ensembleaverage{\Ja(x) \Jbd (0)} \\
	\DL (K) &= \Xint \FT{K}{X}  \ensembleaverage{ \Jbd (0)\Ja(x)}\\
	\SpF (K) &=  \Xint \FT{K}{X}  \ensembleaverage{\TB{\Ja(x), \Jbd (0)}} = \DG(K)-\DL(K) \label{Eq_SpF}
\end{align}
where $\DR$ is the retarded correlator, $\DG$ and $\DL$ are Wightman functions and $\SpF$ is the spectral function. They are not independent of each other. The Kubo-Martin-Schwinger(KMS) relation (in momentum space) connects two Wightman functions  as 
\begin{equation}
	\DG(K) =  e^{\beta k_0} \DL (K) \label{Eq_KMS}~.
\end{equation}
Consequently one can write
\begin{equation}
	\SpF (K)= \DG(K)-\DL(K) = (1 - e^{-\beta k_0}) \DG (K)~.
\end{equation}
Note that using the definition of spectral function in Eq.~\eqref{Eq_SpF} we can write the retarded correlator as
\begin{equation}
	\DR (K) = i \Xint \theta (t) \Pint  \IFT{P}{X} \SpF (P)  \label{Eq_Ret2}
\end{equation}
Now inserting the representation for $\theta$-function
\begin{equation}
	\theta (t) = i\int_{-\infty}^{\infty} \frac{d \omega}{ 2 \pi } \frac{e^{-i \omega t}}{\omega + i \epsilon} 
\end{equation}
in Eq.~\eqref{Eq_Ret2} we get
\begin{align}
	\DR (K) &= - \Pint \int_{-\infty}^{\infty} \frac{d \omega}{ 2 \pi } \frac{1}{\omega + i \epsilon} \int d t     e^{i(k_0- p_0 -\omega ) t}  \int d^3 \vec{x}  e^{-i (\vec{k} - \vec{p})\cdot \vec{x}}  \SpF (p_0 , \vec{p}) \nn \\
	&=  \int \frac{d p_0}{2 \pi}  \int \frac{d \omega}{ 2 \pi } \frac{1}{\omega + i \epsilon}  (2\pi) \delta (k_0- p_0 -\omega) \SpF (p_0 , \vec{k}) \nn  \\
	&=  \int \frac{d p_0}{2 \pi}  \frac{\SpF (p_0 , \vec{k}) }{p_0 - k_0  - i \epsilon} \label{Eq_SRet}
\end{align}
Now let us consider the 11-component of the complex time ordered correlator which is given by
\begin{align}
	D_{\alpha \beta } (K)_{11} &= i \Pint \Xint \FT{(K-P)}{X} \TB{\theta(t) + e^{ - \beta p_0} \theta(-t) } \frac{\SpF (P)}{1 - e^{-\beta p_0}}
	\nn \\
	&= i \Pint \int d^3 x   \frac{\SpF (P)}{1 - e^{-\beta p_0}} \int dt e^{i (k_0 -p_0)t} \TB{ \theta (t) (1 -  e^{-\beta p_0}) + e^{-\beta p_0}} \nn \\
	&= \int \frac{dp_0}{2\pi}     
	\TB{\mathcal{P} \fbr{\frac{1}{k_0 -p_0}  }  + i \pi \coth \fbr{\frac{\beta p_0}{2}} \delta (k_0 - p_0) } \SpF (p_0, \vec{k}) \label{Eq_SD11}
\end{align}
Now comparing Eqs.~\eqref{Eq_SRet} and \eqref{Eq_SD11} we get
\begin{equation}
	\RE D_{\alpha \beta } (K)_{11} = \RE \DR (K)~~~~~~~~~~~\text{and}~~~~~~~~~~~\IM D_{\alpha \beta } (K)_{11} = \coth \fbr{\frac{\beta p_0}{2}}  \IM \DR (K)
\end{equation}


\bibliography{main} 

	\end{document}